\documentclass[9pt,conference]{IEEEtran}
\IEEEoverridecommandlockouts
\usepackage{cite}
\usepackage{amsmath,amssymb,amsfonts}
\usepackage{algorithmic}
\usepackage{graphicx}
\usepackage{textcomp}
\usepackage{xcolor}
\def\BibTeX{{\rm B\kern-.05em{\sc i\kern-.025em b}\kern-.08em
    T\kern-.1667em\lower.7ex\hbox{E}\kern-.125emX}}
\usepackage[all]{nowidow}
\usepackage[english]{babel}
\usepackage{xfrac, xspace}
\usepackage{enumitem}
\usepackage{comment}
\usepackage{balance}
\usepackage{booktabs, tabularx, multirow}
\newcolumntype{Y}{>{\centering\arraybackslash}X}
\newcolumntype{R}{>{\raggedleft\arraybackslash}X}
\newcolumntype{L}{>{\raggedright\arraybackslash}X}

\usepackage{subcaption}
\DeclareCaptionFormat{cap9}{\fontsize{9pt}{10.5pt}\selectfont\textbf{#1#2#3}}
\newcommand{\jmk}[1]{{#1}\xspace}

\newcommand{\NAME}{CiFHER\xspace}
\newcommand{\evk}{$\mathsf{evk}$\xspace}
\newcommand{\evks}{$\mathsf{evk}$s\xspace}

\newcommand{\helr}{\texttt{HELR}\xspace}
\newcommand{\helrsmall}{\texttt{HELR256}\xspace}
\newcommand{\helrlarge}{\texttt{HELR1024}\xspace}
\newcommand{\sorting}{\texttt{Sort}\xspace}
\newcommand{\resnet}{\texttt{ResNet}\xspace}
\newcommand{\bootstrap}{\texttt{Boot}\xspace}
   \makeatletter
\newcommand{\linebreakand}{%
  \end{@IEEEauthorhalign}
  \hfill\mbox{}\par
  \mbox{}\hfill\begin{@IEEEauthorhalign}
}
\makeatother
\begin{document}

\title{CiFHER: A Chiplet-Based FHE Accelerator with a Resizable Structure\\
}

\author{\IEEEauthorblockN{Sangpyo Kim}
\IEEEauthorblockA{\textit{Department of Intelligence and Information} \\
\textit{Seoul National University}\\
Seoul, South Korea\\
vnb987@snu.ac.kr}
\and
\IEEEauthorblockN{Jongmin Kim}
\IEEEauthorblockA{\textit{Interdisciplinary Program in Artificial Intelligence} \\
\textit{Seoul National University}\\
Seoul, South Korea\\
jongmin.kim@snu.ac.kr}
\linebreakand
\IEEEauthorblockN
{Jaeyoung Choi}
\IEEEauthorblockA{\textit{Department of Intelligence and Information} \\
\textit{Seoul National University}\\
Seoul, South Korea\\
sojae0518@snu.ac.kr}
\and
\IEEEauthorblockN{Jung Ho Ahn}
\IEEEauthorblockA{\textit{Department of Intelligence and Information} \\
\textit{Seoul National University}\\
Seoul, South Korea\\
gajh@snu.ac.kr}
}

\maketitle

\begin{abstract}
Fully homomorphic encryption (FHE) is in the spotlight as a definitive solution for privacy, but the high computational overhead of FHE poses a challenge to its practical adoption.
Although prior studies have attempted to design ASIC accelerators to mitigate the overhead, their designs require excessive chip resources (e.g., areas) to contain and process massive data for FHE operations.
We propose \NAME, a chiplet-based FHE accelerator with a resizable structure, to tackle the challenge with a cost-effective multi-chip module (MCM) design.
First, we devise a flexible core architecture whose configuration is adjustable to conform to the global organization of chiplets and design constraints.
Its distinctive feature is a composable functional unit providing varying computational throughput for the number-theoretic transform, the most dominant function in FHE.
Then, we establish generalized data mapping methodologies to minimize the interconnect overhead when organizing the chips into the MCM package in a tiled manner, which becomes a significant bottleneck due to the packaging constraints.
This study demonstrates
that a \NAME package composed of a number of compact chiplets provides performance comparable to state-of-the-art monolithic ASIC accelerators while significantly reducing the package-wide power consumption and manufacturing cost.
\end{abstract}


\section{introduction} \label{sec:introduction}

Privacy is no longer merely a virtue but a critical objective for corporates due to recently established strict data privacy policies, such as the California Consumer Privacy Act (CCPA) and the EU General Data Protection Regulation (GDPR).
For example, Apple's App Tracking Transparency (ATT) policy cost
technology companies \$15.8 billion in 2022~\cite{lotame-2022-att}.

\emph{Fully homomorphic encryption (FHE)} is emerging as an attractive solution providing strong privacy guarantees.
FHE enables an unlimited number of computations directly on encrypted data without decryption; thus, it allows users to safely offload services without disclosing private information.
Numerous studies have shown its applicability to practical privacy-preserving applications, including genomics~\cite{cell-2021-genotype, bigdata-2016-string-genome}, biometrics~\cite{scn-2020-fingerprint, trustcom-2021-biometric}, and machine learning (ML) tasks~\cite{aaai-2019-helr, icml-2022-resnet, access-2022-resnet20, crypto-2018-discretized-nn, cscml-2021-pbs-nn}.

There are two major competing classes of FHE based on the learning with errors (LWE) computational problem: one based on the ring LWE variant (e.g., BGV~\cite{toct-2014-bgv}, BFV~\cite{siam-2014-bfv1, iacr-2012-bfv2}, and CKKS~\cite{asia-2017-ckks} schemes) and the other based on the torus LWE (e.g., FHEW~\cite{eurocrypt-2015-fhew} and TFHE~\cite{jc-2020-tfhe} schemes).
This work focuses on the former class and, specifically, CKKS~\cite{asia-2017-ckks}, which stands out among the state-of-the-art FHE schemes due to its relatively high performance~\cite{iacr-2023-demystify-boot} and the ability to encrypt complex vectors.
Such benefits make CKKS suitable for numerous practical tasks, particularly ML tasks.

Still, alleviating the computational and memory overhead of FHE is a challenging prerequisite for its widespread use as FHE workloads run over 10,000$\times$ slower than their unencrypted counterparts~\cite{isca-2022-craterlake, access-2021-demystify}.
To mitigate the overhead, prior work has attempted hardware acceleration~\cite{wahc-2021-hexl, hpca-2023-tensorfhe, tches-2021-100x, hpca-2023-fab, hpca-2023-poseidon, isca-2022-bts, micro-2021-f1, isca-2022-craterlake, micro-2022-ark}.
In particular, ASIC FHE accelerator proposals~\cite{isca-2022-craterlake, micro-2022-ark, isca-2022-bts} achieved considerable performance enhancements.
For example, ARK~\cite{micro-2022-ark} can perform a CIFAR-10 CNN inference in 0.125 seconds with the ResNet-20 model, which takes 2,271 seconds in a single-threaded CPU~\cite{icml-2022-resnet}.
However, these remarkable speedups owe to the massive chip area usage (373.6--472.3mm\textsuperscript{2}) that enables the deployment of unconventionally large on-chip memory capacity (256--512MB) and substantial amounts of computational logic.

As prior massive FHE accelerator designs are prohibitively costly to realize, we turn to using a \emph{multi-chip module (MCM)}.
The cost of a chip skyrockets with the die area in recent technology nodes due to severe degradation in yield and high design complexity~\cite{whitepaper-2021-hir, micro-2019-simba}.
By splitting a monolithic chip design into several small \emph{chiplets}, an MCM drastically reduces the cost and becomes a scalable solution in the post-Moore era.
Abundant studies~\cite{micro-2019-simba, isca-2021-nnbaton, ieeemicro-2021-manticore, isca-2020-centaur, hpca-2022-spacx, isca-2017-mcmgpu}, including commercial chips~\cite{isscc-2022-sapphire, isca-2021-amd-chiplet, isscc-2020-amdchiplet, ieeemicro-2021-kunpeng}, have demonstrated the efficiency of MCMs in high-performance computing (HPC).

We propose \NAME, a flexible and cost-effective MCM architecture for FHE.
First, \jmk{we propose a core architecture that is resizable at design time.}
Our \emph{composable number-theoretic transform (NTT) unit}, which can provide varying computational throughput depending on the package configuration and technology demands, provides the resizing flexibility, enabling architects to split a monolithic design into multiple cores with a best-fit size for a given die area budget.

Given a chiplet architecture, we construct efficient package configurations of \NAME, focusing on minimizing the high network cost of MCMs.
Due to the technical constraints of a \emph{network on package (NoP)} connecting chiplets, an MCM architecture becomes bottlenecked by the NoP communication among the cores; data communication has been a source of bottleneck even in prior monolithic accelerators, and MCM intensifies the problem.
We address the problem by developing generalized data mapping methodologies optimized for the many-core MCM architecture.
We also devise a \emph{limb duplication} algorithm, which significantly reduces data communication during \emph{base conversion (BConv)}, the second most dominant function in FHE.  

Based on the analysis, we discover balanced combinations of the individual chiplet design, data mapping, and FHE algorithms under various design constraints.
These optimizations enable \NAME to achieve comparable performance to prior monolithic chips; a \NAME package with 16 cores performs an encrypted CIFAR-10 CNN inference using the ResNet-20 model~\cite{icml-2022-resnet} in 0.189 seconds.
Depending on the area budget, \NAME provides various core configurations, ranging from four 47.08mm\textsuperscript{2} core dies to sixty-four 4.28mm\textsuperscript{2} core dies in default settings. 

Overall, this work makes the following key contributions:
\begin{itemize}[noitemsep, leftmargin=*]
    \item We design a flexible chiplet core with a composable NTT unit, which allows the distribution of memory and compute resources across multiple dies while taking advantage of the vector NTT unit.
    \item We introduce generalized data mapping methodologies on a tiled FHE accelerator to resolve the NoP communication bottleneck of MCMs.
    \item We propose limb duplication, an algorithmic optimization tailored to the MCM design, reducing the amount of die-to-die communication, which would have caused significant latency and energy overhead.
\end{itemize}

\section{Background \& Motivation} \label{sec:background}

\subsection{Fully Homomorphic Encryption (FHE)} \label{sec:background:fhe}

\emph{Homomorphic encryption (HE)} is classified into \emph{leveled HE (LHE)} and \emph{fully HE (FHE)} depending on the problem size.
LHE supports only a limited number of operations (ops) and is usually used in conjunction with other cryptographic protocols~\cite{usenixsec-2018-gazelle, security-2022-cheetah}.
In contrast, FHE solely supports an unlimited number of ops.
A defining feature of FHE is \emph{bootstrapping}.
The LWE problem~\cite{jacm-2009-lwe} forms the cryptographic basis for state-of-the-art HE schemes, and relies on the error in encryption for its security.
When performing \emph{HE ops} on encrypted data, the magnitude of errors increases, threatening data integrity unless managed properly.
Bootstrapping reduces the magnitude of errors, thus allowing more HE ops performed.
In a typical use case, a bootstrapping op is inserted after every predetermined number of HE ops is applied to the encrypted data.

\subsection{CKKS FHE Scheme} \label{sec:background:ckks}

We briefly describe the CKKS FHE scheme, the main target of this paper.
During CKKS encryption, a (vector) \emph{message} composed of $n$ numbers is first \emph{packed} into a degree-($N-1$) integer polynomial referred to as \emph{plaintext}, which is an element of a cyclotomic polynomial ring $\mathcal{R}_Q = \mathbb{Z}_Q[X]/(X^N + 1)$ with the \emph{degree} $N \geq 2n$ and the \emph{modulus} $Q$.
All integer ops are performed modulo-$Q$ in $\mathcal{R}_Q$.
A plaintext is \emph{encrypted} into a \emph{ciphertext} composed of two polynomials in $\mathcal{R}_Q$.
We denote polynomials with bold characters (e.g., $\mathbf{v}$), and ciphertexts encrypting polynomials with brackets ($[\mathbf{v}]$).
Then, the encryption of $\mathbf{v}$ can be formulated as 
$[\mathbf{v}] = (\mathbf{v}_1, \mathbf{v}_2) \in \mathcal{R}_Q^2$\text{.}

A high $N$ value around $2^{15}$--$2^{17}$ and $Q$ as large as $2^{1500}$ are utilized to meet the security constraints of FHE~\cite{2021-standard, wahc-2019-curtis}.
To handle such a large $Q$, \emph{residue number system (RNS)} is utilized~\cite{sac-2018-frns-ckks}.
$Q$ is set to the product of $L$ number of small primes ($q_i$'s), such that $Q=\prod_{i=1}^L q_i$.
Then, each coefficient $c$ of a polynomial is represented by an $L$-tuple of small \emph{residues} $(c \text{ mod } q_1, \cdots, c \text{ mod } q_L)$.
Each row of the matrix corresponding to the prime $q_i$ is referred to as \emph{limb}, which can be regarded as a polynomial in $\mathcal{R}_{q_i} = \mathbb{Z}_{q_i}[X]/(X^N + 1)$.
Using RNS, modulo-$Q$ ops between polynomials having large coefficients are replaced by a set of $L$ pairwise modulo-$q_i$ ops between limbs having small word-sized coefficients.
Also, the modulus can change over HE ops such that $Q_\ell = \prod_{i=1}^\ell q_i$ is used at a particular moment.
As a result, a polynomial at that moment is represented as an $\ell \times N$ matrix of residues.

In a client-server scenario of FHE, the server performs \emph{HE ops} to manipulate the message encrypted in a ciphertext.
Mult/add between two ciphertexts ($\mathtt{HMult}$/$\mathtt{HAdd}$) or between a ciphertext and a plaintext ($\mathtt{PMult}$/$\mathtt{PAdd}$) are some of the most frequently used HE ops.
Also, it is often required to change the data order inside the message vector composed of $n$ numbers.
Such data reordering can be performed with an $\mathtt{HRot}$ op, which evaluates the cyclic rotation of the message encrypted inside a ciphertext. 
Among these HE ops, $\mathtt{HMult}$ and $\mathtt{HRot}$ are the most computationally expensive due to complex \emph{key-switching} procedures involved in them.
Key-switching is a procedure to multiply a polynomial with \emph{evaluation keys} (\evks), which are special types of ciphertexts.

\subsection{Primary Functions of CKKS} \label{sec:background:primary_functions}

We defer a more detailed explanation of each HE op to prior work~\cite{asia-2017-ckks, sac-2018-frns-ckks, rsa-2020-better, eurocrypt-2021-efficient} and instead break down HE ops into \emph{primary functions} to discuss their computational aspects.

\noindent \textbf{Number-theoretic transform (NTT)}:
To reduce the complexity of a polynomial mult, NTT is utilized.
A polynomial mult in $\mathcal{R}_{q_i}$ is equivalent to a negacyclic convolution between the coefficients.
NTT, a type of discrete Fourier transform, and its inverse (iNTT) are defined
for each $q_i$.
(i)NTT reduces the overall complexity of the convolution into $O(N \log N)$ for each limb by adopting a fast Fourier transform (FFT) algorithm~\cite{cooley-tukey}.

\noindent \textbf{Base conversion (BConv)}:
The modulus $Q$ changes frequently over HE ops, which requires BConv to be performed on polynomials.
For example, for a polynomial $\mathbf{v} \in \mathcal{R}_{Q_\ell}$, we often need to change its modulus to an \emph{auxiliary modulus} $P=\sum_{i=1}^K p_i$.
For this situation, 96\% of BConv computation is spent on a matrix-matrix mult between a pre-computed \emph{BConv table} (a $K \times \ell$ matrix) and $\mathbf{v}$ (an $\ell \times N$ matrix)~\cite{micro-2022-ark}.
We obtain a new $\mathbf{v} \in \mathcal{R}_P$ through BConv.

\noindent \textbf{Automorphism ($\phi_r$)}:
During $\mathtt{HRot}$, automorphism, which is a data permutation function, is performed.
The $i$-th coefficient of a polynomial is mapped to the ($i\cdot5^r \text{ mod } N$)-th position.

\noindent \textbf{Element-wise or constant functions}:
Other than the aforementioned primary functions, we only perform element-wise mult/add or constant mult/add on polynomials for HE ops.

FHE computations are extremely expensive, mainly attributed to key-switching, which involves multiple (i)NTTs and BConvs.
(i)NTT and BConv account for more than 80\% of the total computations in CKKS~\cite{isca-2022-bts, micro-2022-ark}.
Thus, enhancing (i)NTT and BConv performance is essential to FHE acceleration.

\subsection{Previous ASIC FHE Accelerators} \label{sec:background:prior}

F1~\cite{micro-2021-f1} is the first ASIC accelerator that reports bootstrapping performance.
However, F1 targets small parameters of $N\!=\!2^{14}$ and $L\!=\!16$, with which it only partially supports bootstrapping.
Subsequent accelerator proposals, BTS~\cite{isca-2022-bts}, CraterLake~\cite{isca-2022-craterlake}, and ARK~\cite{micro-2022-ark}, fully support bootstrapping by targeting larger parameters of $N=2^{16}$--$2^{17}$ and $L=24$--$60$.
BTS places 2,048 processing elements (PEs) in a grid and connects them by global wires that span the whole chip.
Each PE contains a \emph{butterfly unit} for (i)NTT, a multiply-accumulate unit for BConv, and a multiplier and an adder for element-wise functions.
A butterfly unit is capable of performing the \emph{butterfly ops} required in (i)NTT, which are shown in Eq.~\ref{eq:butterfly}.
ARK and CraterLake develop upon F1's design of placing $\sqrt{N}$ parallel \emph{lanes} in a cluster (CraterLake refers to it as lane group) that feeds data into massive \emph{vector NTT units (NTTUs)}.
A vector NTTU contains $\frac{1}{2}\sqrt{N}\log N$ butterfly units, placed in an organization directly reflecting the four-step FFT dataflow~\cite{sc-1989-fft4}.
ARK deploys four NTTUs, and CraterLake 16.
\begin{equation} \label{eq:butterfly}
\begin{split}
\text{I. }&\mathtt{Butterfly}_\text{NTT}(a, b, w) = (a + b\cdot w, \ a - b \cdot w) \\
\text{II. }&\mathtt{Butterfly}_\text{iNTT}(a, b, w) = (a + b, \ (a - b) \cdot w) 
\end{split}
\end{equation}

To tackle the increased number of computations and size of the working set when using large parameters, CraterLake, BTS, and ARK all rely on enormous amounts of computational logic and on-chip memory capacity; the latter even reaches 256--512MB.
These designs end up using massive chip resources, utilizing a monolithic die sized 373.6--472.3mm\textsuperscript{2} and dissipating up to 317W of power.

\subsection{Multi-Chip Module (MCM)} \label{sec:background:mcm}

MCM is a promising solution for scaling chips in the post-Moore era.
By organizing a module with multiple small dies, referred to as \emph{chiplets}, and connecting them using advanced packaging technologies providing high bandwidth on-package interconnects~\cite{ted-2017-cowos, ectc-2016-emib, iedm-2022-focos}, MCMs reduce the area of an individual die.
Because the design and fabrication cost of a die substantially drops as the die area is reduced~\cite{whitepaper-2021-hir, micro-2019-simba}, MCMs can greatly reduce the total cost.
With yield and technology scaling declining over chip fabrication generations, MCMs have become one of the few left options for scaling systems.
Recent Intel~\cite{isscc-2022-sapphire} and AMD~\cite{isca-2021-amd-chiplet, isscc-2020-amdchiplet} CPUs are notable examples of MCMs; multiple chiplets containing several CPU cores and cache memory are integrated into a single package using EMIB~\cite{ectc-2016-emib} or CoWoS~\cite{ted-2017-cowos} packaging technologies.
Also, numerous domain-specific architectures utilizing MCMs have been proposed~\cite{micro-2019-simba, isca-2021-nnbaton, isca-2020-centaur, hpca-2022-spacx, ieeemicro-2023-dojo}.
Moreover, \jmk{MCMs have been realized at various area granularity}, ranging from AMD's high-end 74mm\textsuperscript{2} die~\cite{isscc-2020-amdchiplet} to  6mm\textsuperscript{2} die of Simba~\cite{micro-2019-simba}.

These recent trends motivate us to design an MCM architecture for cost-effective FHE acceleration.
By using MCMs, we can enjoy the aforementioned benefits coming from a significant reduction in individual die areas.
Also, this approach is highly scalable as we can reuse the same design for each chiplet and place multiple chiplets to facilitate developing various MCM FHE accelerators with different computational throughput and bandwidth required by the technology constraints and the purpose of use.
Prior ASIC FHE accelerator proposals include high bandwidth memories (HBMs), already assuming the use of advanced packaging for connecting HBMs to the chip.
The same packaging can be utilized to connect multiple chiplets with little additional cost~\cite{micro-2015-disintegration}.
We utilize the open specification of Universal Chiplet Interconnect Express (UCIe)~\cite{whitepaper-2022-ucie} to design and evaluate the MCM FHE accelerator.

\begin{figure*}[t]
\centering
  \includegraphics[height=2.4in]{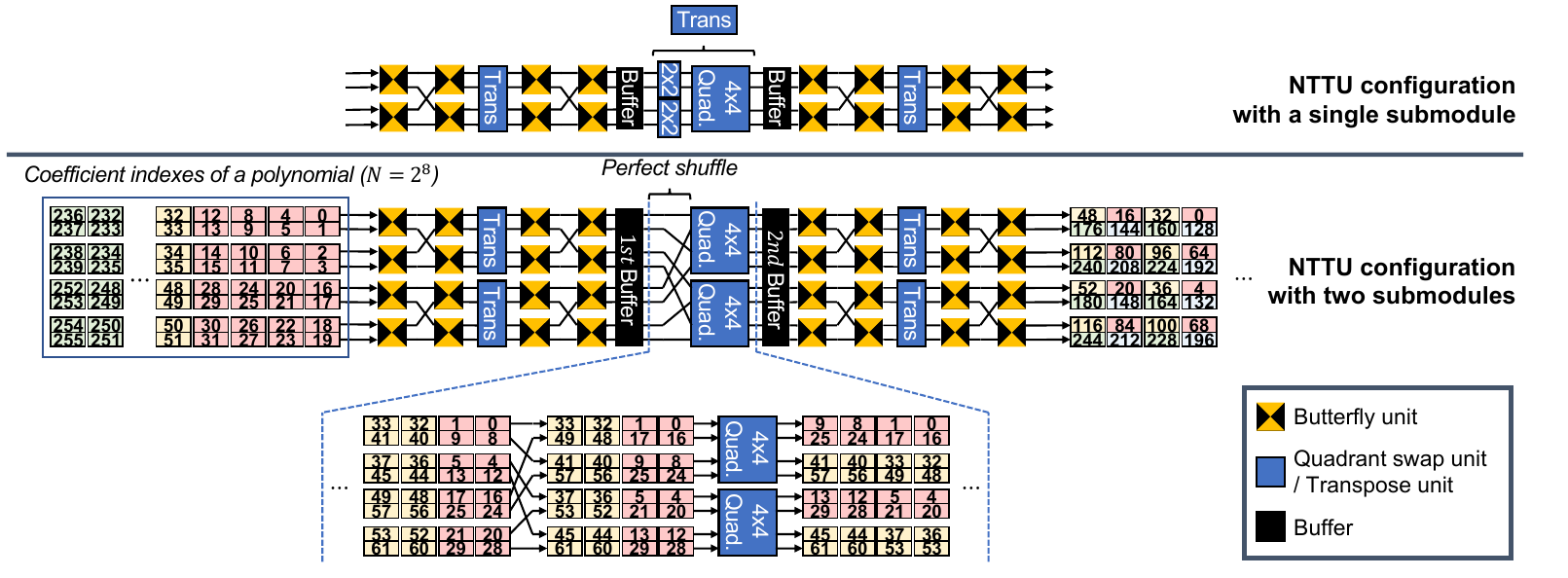}
\caption{Exemplar configurations of a composable NTT unit, simplified to  $N=2^8$. A submodule, the smallest unit occupying $\sqrt[4]{N}$ lanes, is shown above. A two-submodule configuration and its NTT process for a length-$N$ polynomial are shown below.}
    \label{fig:nttu}
    \vspace{-0.05in}
\end{figure*}

\begin{figure}[t]
\centering
  \includegraphics[height=1.6in]{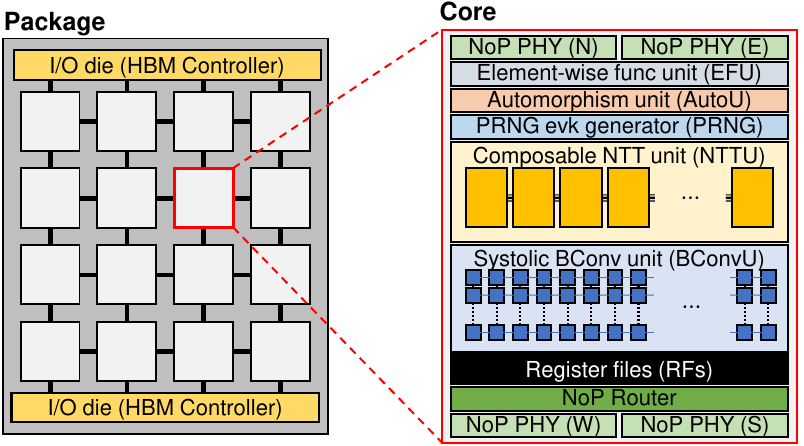}
  
\caption{
The \NAME package consists of multiple core chiplets and two I/O dies handling HBM.
A core comprises functional units (FUs), register files (RFs), and networking components.}
    \label{fig:floorplan}
    \vspace{-0.05in}
\end{figure}

\section{\NAME Microarchitecture}

To reduce the high cost of a monolithic chip design, we propose \NAME, an MCM architecture for FHE acceleration.
\NAME has a tiled architecture with multiple chiplet \emph{cores} connected by a mesh NoP.
Fig.~\ref{fig:floorplan} depicts the package and core organization of \NAME.
The main challenge in designing an MCM architecture lies in reducing the NoP pressure.
As deployable bandwidth and topology are heavily restricted by the technology constraints of the NoP, we devise novel data mapping methodologies (\S\ref{sec:contribution-1}) and algorithms (\S\ref{sec:contribution-2}) addressing the high latency and low bandwidth of NoP.
Nonetheless, prior to determining the configuration of the MCM package, we will first architect the internals of a chiplet core. 

\subsection{Core Organization}

Prior accelerators either utilize little (PE in BTS) or big (cluster or lane group in F1, ARK, and CraterLake) cores without providing much flexibility in design.
In contrast, we design the cores of \NAME to have varying computational throughput, which can be adjusted at design time depending on the package configuration.

A core comprises a \emph{composable NTTU}, a \emph{systolic BConv unit (BConvU)}, an \emph{automorphism unit (AutoU)}, an \emph{element-wise function unit (EFU)}, a \emph{PRNG \evk generator (PRNG)}, a router and PHYs for NoP, 
and register files (RFs).
We adopt the vector architecture utilized in F1, CraterLake, and ARK, where we place multiple parallel lanes in a core.
Each lane has a dedicated RF space not accessible by the other lanes, which enables maximizing parallelism with low costs for synchronization and arbitration.
The vector architecture has shown effectiveness in reducing the communication overhead and the RF bandwidth pressure, the major bottlenecks in FHE.

By effectively combining components from different vector accelerators with proper modifications, \NAME provides an improved core design for FHE accelerators.
In particular, we adopt the distributed RF, systolic BConvU, and AutoU structures from ARK, and adopt PRNG from CraterLake.
A systolic BConvU is composed of multiple multiply-accumulate (MAC) units, forming an output-stationary systolic array.

To achieve a flexible core design, we aim to adjust configurations, particularly the number of lanes. While most functional units (FUs) readily adapt to lane count and throughput goals, the preceding vector NTTU presents a challenge. This critical vector core component is a substantial computing unit with its 2,048 butterfly units locked into a fixed organization. 

The inflexibility of previous large-scale NTTU designs severely limits our design options, ultimately harming hardware efficiency.
To manage the on-chip memory bandwidth demands of NTT computation, prior studies rely on deeply pipelined NTTU designs.
These treat a length-N polynomial as an $\sqrt{N} \times \sqrt{N}$ matrix, performing the NTT in four steps: column-wise $\sqrt{N}$-point NTT, twisting, transposition, and row-wise $\sqrt{N}$-point NTT.
With $N$ reaching $2^{16}\!-\!2^{17}$, this vector NTTU becomes bulky and expensive.
Worse, its single-configuration design forces us to use large chiplets to accommodate it, regardless of the actual computational throughput needed.
This leads to underutilized NTTUs, creating an imbalanced and inefficient design for MCMs.

\subsection{Composable NTT Unit}

We devise a composable NTTU supporting various configurations with the number of lanes adjustable from 16 to 256.
We also regard a length-$N$ polynomial as a $\sqrt{N}\times\sqrt{N}$ matrix and perform $\sqrt{N}$-point (i)NTT along the rows and then along the columns in sequence, but additionally apply the four-step FFT dataflow~\cite{sc-1989-fft4} to each of the row- and column-direction (i)NTTs.
We obtain a compact NTTU spanning $\sqrt[4]{N} = 16$ lanes shown in Fig.~\ref{fig:nttu} (simplified to $N=2^8$), which we refer to as the \emph{submodule} of our NTTU.

Up to 16 submodules can be stacked to deliver higher computational throughput for (i)NTT, providing flexibility for our design space exploration.
We devise a method to combine multiple submodules to have them collaboratively perform (i)NTT on a limb by performing a perfect shuffle~\cite{tc-1971-perfect-shuffle} between the submodules.
The perfect shuffle enables the rearrangement of dispersed data across various submodules by transforming it into a sequence that can be effectively transposed using subsequent quadrant swap units.
An exemplar two-submodule NTTU configuration for $N\!=\!2^8$ performing NTT is shown in Fig.~\ref{fig:nttu}.
We regard a length-$2^8$ polynomial as a $16\!\times\!16$ matrix.
The first half of a submodule performs a four-step NTT on a row (data indices: 0--15/16--32).
It starts from stride-1 butterfly ops and ends with stride-8 ops to perform 16-point NTT on a row.
Then the buffering and shuffling steps in the middle redistribute the data between different submodules so that each submodule holds data elements with stride 16 in the second buffer; i.e., each submodule holds the elements of a column.
Finally, the second half of a submodule performs a four-step NTT on a column (data indices: 0--240/4--244, stride 16) by starting from stride-16 butterfly ops and ending with stride-128 ops to perform 16-point NTT on a column.

Our composable NTTU also allows up to $\sqrt[4]{N}=16$ cores to perform (i)NTT of a limb together.
In this case, data exchange between the collaborating cores occurs after the buffering and shuffling steps in the case of NTT.

\setlength{\tabcolsep}{1.8pt}
\begin{table}
\small
\centering
\caption{Target parameters of \NAME.} \label{tab:parameter}
\begin{tabularx}{0.995\columnwidth}{l|YYYYYc}
   \toprule
   Param. & $N$ & $L$ & $K$ & $Q$ & $P$ & Security \\
   Descr. & Degree & \# of $q_i$ & \# of $p_i$ & $\prod_{i=1}^L q_i$ & $\prod_{i=1}^K p_i$ & (bits) \\
   \midrule
   Value     & $2^{16}$   & $\leq48$ & 12 & $\leq2^{1218}$ & $2^{336}$ & $\geq128$~\cite{eurocrypt-2021-efficient} \\
   \bottomrule
\end{tabularx}
\vspace{-0.05in}
\end{table}
\setlength{\tabcolsep}{6pt}

\subsection{Word Length \& Logic Optimization}

We choose 32 bits as the word length to use in \NAME as it is beneficial to use short word lengths for the efficiency of logical operations.
While BTS and ARK utilize a word length of 64 bits, which has been traditionally used in CKKS libraries~\cite{github-lattigo, github-heaan, fcds-2017-seal}, F1 (32 bits), CraterLake (28 bits), and FPGA FHE accelerators (FAB~\cite{hpca-2023-fab} (54 bits) and Poseidon~\cite{hpca-2023-poseidon} (32 bits)) utilize shorter word lengths.
Despite using a relatively short word length, we maintain high precision by the methods in \cite{wahc-2023-wordsize} to utilize high scales ranging from $2^{47}$ to $2^{55}$.
Table~\ref{tab:parameter} tabulates our target parameters.

We utilize the word-level Montgomery reduction circuit proposed in \cite{dsd-2019-mert} with further enhancements using the signed Montgomery reduction algorithm in \cite{iacr-2018-signed-mont}.
The circuit is utilized for all modular reduction circuits in \NAME, including EFUs, NTTUs, PRNGs, and BConvUs.

\noindent\textbf{Systolic BConvU:} By using a 32-bit word length, more limbs are utilized (higher $L$ and $K$) than 64-bit ARK.
Thus, to provide higher word-level throughput for BConv, we use a longer $1\!\times\!12$ BConv configuration per lane by default.

\noindent\textbf{EFU:} Inside an EFU, we place multiple modular multipliers, modular adders, and element-wise op circuits for specific purposes required in FHE (e.g., double-word accumulation and reduction).
An EFU can perform compound element-wise ops to reduce the pressure on RFs.

\section{Data mapping methodology} \label{sec:contribution-1}

\subsection{Tiled Architecture}
\label{sec:contribution-1:tiled}

The technology constraints of NoP enforce \NAME to be a tiled architecture connected by a mesh network.
Due to the short channel reach of an advanced interface~\cite{whitepaper-2022-ucie}, a complex network topology that requires connections between distant nodes is less suitable for MCMs.
Thus, we fix the topology to 2D mesh (see Fig.~\ref{fig:floorplan}).
Also, due to the high cost of NoP communication, we set the default bisection bandwidth of the entire package to 2TB/s, several to an order of magnitude lower than that of CraterLake (29TB/s) or ARK (8TB/s).

Prior FHE accelerators co-design the data mapping method with the network on chip (NoC) organization.
They design complex NoC structures using direct connections between distant nodes with high bandwidth and uniform communication time.
However, in an MCM architecture with a mesh network, communication latency between nodes is non-uniform, as distant nodes are separated by multiple hops.
Lagging data from distant nodes causes others to wait idly, degrading the overall performance~\cite{micro-2019-simba}.
The low bandwidth of NoP exacerbates this problem.
Therefore, we delve into data mapping methodologies to reduce the pressure of the network.

\subsection{Combining Data Mapping Methods}
\label{sec:contribution-1:combining}

Prior data mapping methods can be classified into \emph{coefficient scattering} and \emph{limb scattering}. 
Coefficient scattering is a method of distributing $N$ coefficients of each limb equally to the entire computing units. 
By contrast, in limb scattering, each prime and the corresponding limbs are allocated to a specific computing unit.
BTS and CraterLake use coefficient scattering, whereas F1 and ARK mostly use limb scattering.
ARK temporarily switches to coefficient scattering during BConv, which we detail in \S\ref{sec:contribution-2:duplication}.

Due to the conflicting data access patterns of HE ops, core-to-core communication is inevitable regardless of the data mapping method.
(i)NTT and BConv, the two most dominant primitives, require different data mapping due to their data access patterns. 
For a polynomial, expressed as an $\ell\times N$ matrix, (i)NTT can be separately applied to each of $\ell$ limbs, favoring limb scattering.
By contrast, as BConv is mostly a matrix-matrix mult with a $K\times \ell$ BConv table (see \S\ref{sec:background:primary_functions}), data access is performed along the columns of a polynomial and makes coefficient scattering a better option.
Thus, data exchange between cores is required either during (i)NTT (coefficient scattering) or BConv (limb scattering).

In an accelerator with many cores, both data mapping methods have limitations.
As the number of limbs ($\ell$) vary over HE ops, it is difficult to distribute the limbs equally to cores without causing fragmentation issues in limb scattering.
A core will be assigned fewer limbs than others depending on $\ell$ and wait idly for others to finish processing.
In contrast, coefficient scattering is free from fragmentation issues because the degree of a polynomial ($N$) does not change.
However, coefficient scattering has a scaling limit due to quadratic growth in the number of transferred packets with increasing cores.
For example, during (i)NTT with $c$ cores, coefficient scattering results in all-to-all data exchange between the cores, requiring a total of $c(c-1)$ packets to be transferred.
These issues become more severe as more cores participate in the distribution.
Also, with more cores, additional performance degradation from long tail latency in lagging data occurs. 

Therefore, in an MCM architecture with many cores and restrained NoP bandwidth, the combined use of limb scattering and coefficient scattering enhances performance, even with the increased total amount of transferred data due to performing data exchanges for both (i)NTT and BConv.
We partition a polynomial in both limb and coefficient directions by dividing cores into \emph{limb clusters} and \emph{coefficient clusters}. 
A limb cluster is a set of cores each holding an evenly distributed subset of the coefficients of a limb.
By contrast, the cores in a coefficient cluster partition the residues of a coefficient corresponding to different primes.
Then, (i)NTT or BConv only incurs data exchange within each limb cluster or coefficient cluster, reducing the number of cores participating in the exchange.
Thus, combining two data mapping methods suffers less from the aforementioned issues, leading to performance enhancement.

\begin{figure}[t]
\centering
\begin{subcaptiongroup}
\includegraphics[height=1.42in]{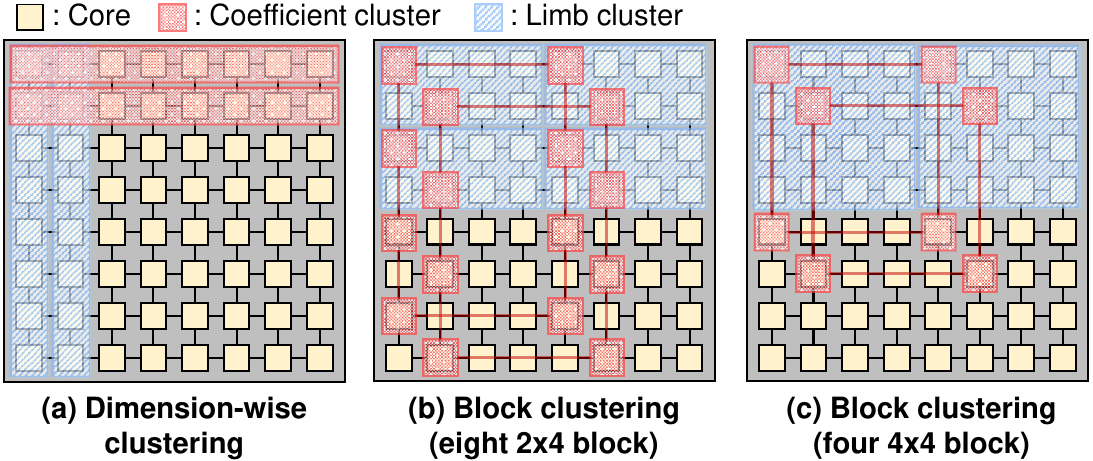}
\phantomcaption\label{fig:mapping:a} 
\end{subcaptiongroup}
\begin{subcaptiongroup} 
\phantomcaption\label{fig:mapping:b}
\end{subcaptiongroup}
\begin{subcaptiongroup} 
\phantomcaption\label{fig:mapping:c}
\end{subcaptiongroup}
\vspace{-0.13in}
  \caption{Generalized data mapping methods of \NAME with 64 cores: \subref{fig:mapping:a} dimension-wise clustering and block clustering for two different exemplar block sizes of \subref{fig:mapping:b} $2\times4$ and \subref{fig:mapping:c} $4\times4$.}
  \label{fig:mapping}
  \vspace{-0.13in}
\end{figure}

\subsection{Generalized Data Mapping for a Tiled FHE Accelerator}
\label{sec:contribution-1:generalized}

We introduce two generalized data mapping methods that combine coefficient scattering and limb scattering for mapping HE ops to a tiled architecture: \emph{dimension-wise clustering} and \emph{block clustering}.
Dimension-wise clustering binds the cores on the same vertical line into a limb cluster and the cores on the same horizontal line into a coefficient cluster (or in the oppositice directions) on a mesh network.
Fig.~\ref{fig:mapping:a} shows an example of dimension-wise clustering on an $8\times8$ mesh.
This distribution method has the advantage that each data exchange process of (i)NTT and BConv can be performed without interference because data movement occurs in different axial directions.
However, the configuration is fixed for a given mesh shape, potentially sharing the same limitations of prior mapping methods when the mesh shape is skewed.

Block clustering is a more generalized mapping method that places limb clusters in the form of blocks with an arbitrary size, and forms coefficient clusters with the cores in the same position within each block.
Fig.~\ref{fig:mapping:b} and Fig.~\ref{fig:mapping:c} show configurations dividing an $8\times8$ mesh into eight $2\times4$ or four $4\times4$ limb cluster blocks.
Unlike dimension-wise clustering, various block clustering configurations are possible by changing the size of a block.
We refer to the configuration on a $d_x \times d_y$ mesh using a $b_h \times b_w $ block size as $d_x\times d_y$-BK-$b_h\times b_w$.
Dimension-wise clustering is simply denoted as $d_x \times d_y$-DW, which is indeed a special variant of block clustering; e.g., Fig.~\ref{fig:mapping:a} can be expressed as $8\times8$-BK-$8\times1$.
Also, limb scattering and coefficient scattering can be expressed as $d_x \times d_y$-BK-$1 \times 1$ and $d_x \times d_y$-BK-$d_x \times d_y$, respectively.

Determining the data mapping method of \NAME is challenging as it requires considering various factors that have conflicting effects on its performance.
In addition to the trade-offs discussed in \S\ref{sec:contribution-1:combining}, generalized data mapping methods reduce the maximum number of hops a packet travels by restricting its destination to adjacent cores within a cluster.
Also, it is challenging to balance the limb clusters and coefficient clusters; an unbalanced configuration would result in fragmentation or severe performance degradation as either (i)NTT or BConv ops must wait for the other to finish due to the dependency in common CKKS use cases.
\jmk{We account for these challenges by developing an extensive simulation tool for \NAME, which we use for the evaluation in \S\ref{sec:evaluation}.}
\section{Algorithmic optimizations} \label{sec:contribution-2}

\subsection{Limb Duplication for BConv} \label{sec:contribution-2:duplication}

\begin{figure}[t]
    \centering
    \includegraphics[height=1.6in]{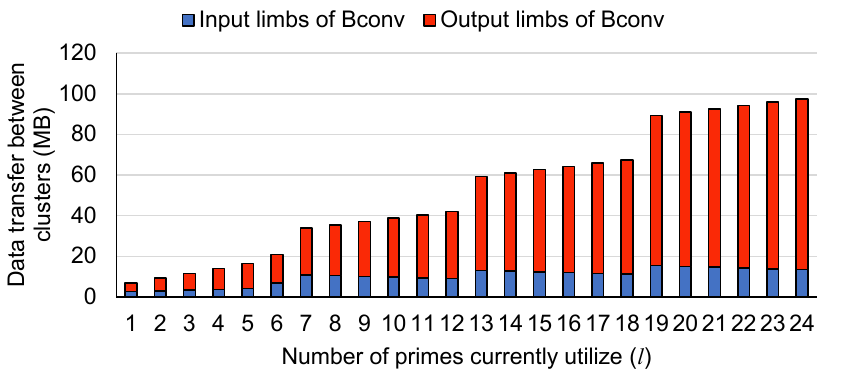}
    \caption{The amount of data transfer during key-switching using the method of ARK~\cite{micro-2022-ark} under different $\ell$ conditions.}
    \label{fig:input_output_limb}
    \vspace{-0.05in}
\end{figure}

We propose a novel \emph{limb duplication} algorithm, which duplicates the input limbs of BConv to reduce the amount of data communication caused by the redistribution of the output limbs during BConv.
BConv requires gathering the residues from different cores.
F1 uses parameters which allows performing BConv in parallel for each of the $\ell$ limbs without redistribution of input data ($\ell$ limbs) before BConv~\cite{fcds-2017-seal}.
However, F1's method produces massive $\ell^2$ output limbs, which need to be redistributed among cores.
More recently proposed ARK adopts parameters that make BConv produce fewer limbs (roughly $\ell\cdot \beta$ limbs for a small integer $\beta$), which also significantly reduces the amount of computation~\cite{rsa-2020-better}.
However, as it becomes harder to parallelize BConv, ARK switches to coefficient scattering prior to BConv and switches back to limb scattering after BConv, inducing data redistribution of both input and output data of BConv.
Fig.~\ref{fig:input_output_limb} shows the estimated amount of data transfer required for key-switching depending on $\ell$ (the number of primes) when using ARK's method.
As the number of input limbs is usually smaller than the number of output limbs, the output limbs of BConv account for most of the data transfer regardless of $\ell$.

To eliminate the data transfer required for the redistribution of generated limbs, limb duplication duplicates input limbs and broadcasts them to all the cores in a coefficient cluster so that each core holds the entire polynomial ($\ell\times N$ matrix when only considering limb scattering). 
Then, each core performs BConv with a piece of the BConv table ($K \times\ell$ matrix); e.g., a core in charge of the primes $p_1, p_5$, and $p_9$ would only multiply the 1\textsuperscript{st}, 5\textsuperscript{th}, and 9\textsuperscript{th} rows of the BConv table with the polynomial to obtain BConv results.
As the output limbs already belong to the right owner, data redistribution after BConv is unnecessary.

However, limb duplication requires additional data transfer for broadcasting input limbs.
The overhead of broadcasting varies depending on the ratio of the number of input limbs and the number of output limbs and the data mapping methodology.
We can formulate the reduction in the number of transferred limbs compared to the ARK's method as in Eq.~\ref{eq:theoretical_benefit}.
$\mathtt{broadcast}_\text{overhead}$ denotes the ratio of data transfer required for broadcasting compared to that required for even distribution and has a value greater than one.
\begin{equation} \label{eq:theoretical_benefit}
\begin{split}
\mathtt{\#limbs}_\text{output} - \mathtt{\#limbs}_\text{input}\times (\mathtt{broadcast}_\text{overhead} - 1)
\end{split}
\end{equation}

Eq.~\ref{eq:theoretical_benefit} must be larger than zero to make limb duplication beneficial. 
We selectively use limb duplication only when this condition is satisfied for each BConv.
As there are much more output limbs than input limbs in general, limb duplication is usually beneficial for the reduction of communication.

\subsection{An Eclectic Approach to Prior Algorithms}

We adopt the state-of-the-art CKKS algorithms mostly from an FHE CPU library, Lattigo~\cite{github-lattigo}.
We combine the bootstrapping algorithms from \cite{eurocrypt-2018-slimboot, acns-2022-sparseboot}.
There are also accelerator-specific algorithms, which offers trade-offs between computation and off-chip memory access.
They target \evks and plaintext polynomials, which are loaded from the off-chip memory and are large in size, possibly inducing the off-chip memory bandwidth bottleneck.
We explain how such algorithms can (or cannot) be applied to \NAME.

\noindent\textbf{Minimum key-switching:}
%
To reduce the memory access for \evks, ARK builds on the algorithm proposed in \cite{crypto-2018-linear} to suggest the minimum key-switching algorithm for HRot ops.
Most of the \evks are for $\mathtt{HRot}$ ops, which require a different $\mathsf{evk}_r$ for each rotation amount ($r$).
The algorithm transforms a common computational pattern of sequential HRot ops with different rotation amounts forming an arithmetic progression into a recursive sequence of HRot ops with the same rotation amount equal to the common difference of the progression.
Then, only one $\mathsf{evk}_r$ is required for the whole sequence.
\NAME also adopts minimum key-switching as it is crucial for performance to reduce the off-chip memory access for \evks.

\noindent\textbf{On-the-fly limb extension:}
Plaintexts (polynomials) required for $\mathtt{PMult}$ ops also account for a considerable portion of off-chip memory access.
For example, bootstrapping requires several GBs of plaintexts.
On-the-fly limb extension compresses each plaintext to occupy only one limb per polynomial and generates the rest at runtime.
However, the runtime extension involves NTT, incurring expensive NoP data transfer.
As the overall cost for runtime extension exceeds the benefits coming from the reduced off-chip memory access in an MCM architecture, we do not use this algorithm in \NAME.

\noindent\textbf{PRNG \evk generation:}
We adopt the PRNG \evk generation from CraterLake.
Half of polynomials composing \evks are purely random.
Therefore, by deploying deterministic PRNG circuits inside a chip, they can be generated on-the-fly using a short seed, reducing the off-chip memory access for \evks to half~\cite{isca-2022-craterlake}.
As \evks can be generated within each core in parallel, no additional overhead is required for MCMs.
\section{Evaluation} \label{sec:evaluation}

\setlength{\tabcolsep}{1.5pt}
\renewcommand{\arraystretch}{1.05}
\begin{table}[t] \centering
\caption{Area breakdown of the default configurations of \NAME with varying numbers of cores.}
\label{tab:area}
\small
\begin{tabularx}{0.99\columnwidth}{r|RRRRR}
\toprule
& \multicolumn{5}{c}{\textbf{Area / core (mm\textsuperscript{2})}} \\
\hline
\multirow{2}{*}{Configuration} & \multicolumn{1}{c}{4 core} & \multicolumn{1}{c}{8 core} & \multicolumn{1}{c}{16 core} & \multicolumn{1}{c}{32 core} & \multicolumn{1}{c}{64 core} \\
& \multicolumn{1}{c}{($2\times2$)} & \multicolumn{1}{c}{($2\times4$)} & \multicolumn{1}{c}{($4\times4$)} & \multicolumn{1}{c}{($4\times8$)} & \multicolumn{1}{c}{($8\times8$)} \\
\hline
Register files & 31.71 & 15.86 & 7.93 & 3.96 & 1.98 \\
NTTU & 3.75 & 1.82 & 0.88 & 0.43 & 0.22 \\
BConvU & 1.07 & 0.65 & 0.44 & 0.34 & 0.28 \\
EFU & 0.72 & 0.36 & 0.18 & 0.09 & 0.05 \\
AutoU & 2.32 & 0.58 & 0.14 & 0.04 & 0.01 \\
PRNG & 0.71 & 0.36 & 0.18 & 0.09 & 0.04 \\
Router/PHY & 6.80 & 3.40 & 3.40 & 1.70 & 1.70 \\
\hline
Total & 47.08 & 23.02 & 13.15 & 6.65 & 4.28 \\
\midrule
 & \multicolumn{5}{c}{\textbf{Package area  (mm\textsuperscript{2})}} \\
\hline
Cores & 188.32 & 184.13 & 210.43 & 212.75 & 273.87 \\
\hline
I/O dies & \multicolumn{5}{c}{36.71} \\
\hline
Total & 225.04 & 220.84 & 247.14 & 249.46 & 310.59 \\
\bottomrule                     
\end{tabularx}
\vspace{-0.13in}
\end{table}
\renewcommand{\arraystretch}{1.0}
\setlength{\tabcolsep}{6pt}

\begin{figure*}[t]
\centering
\begin{subfigure}{.189\textwidth}
  \centering
  \includegraphics[height=1.17in]{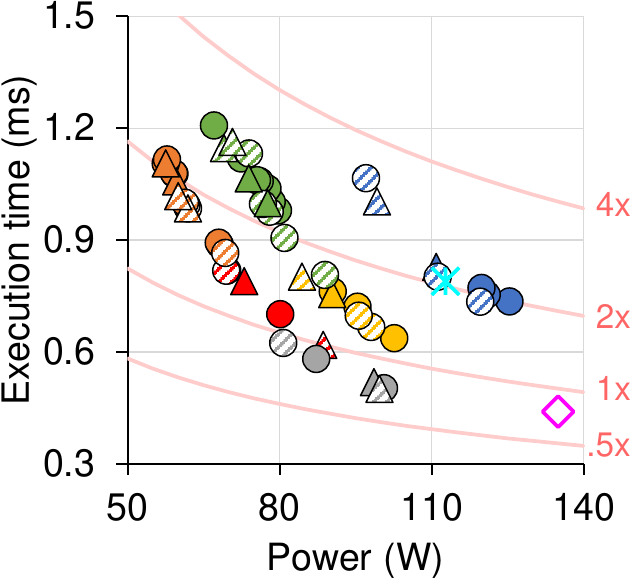}
  \caption{\bootstrap}
  \label{fig:pareto-boot}
\end{subfigure}%
\begin{subfigure}{.182\textwidth}
  \centering
  \includegraphics[height=1.17in]{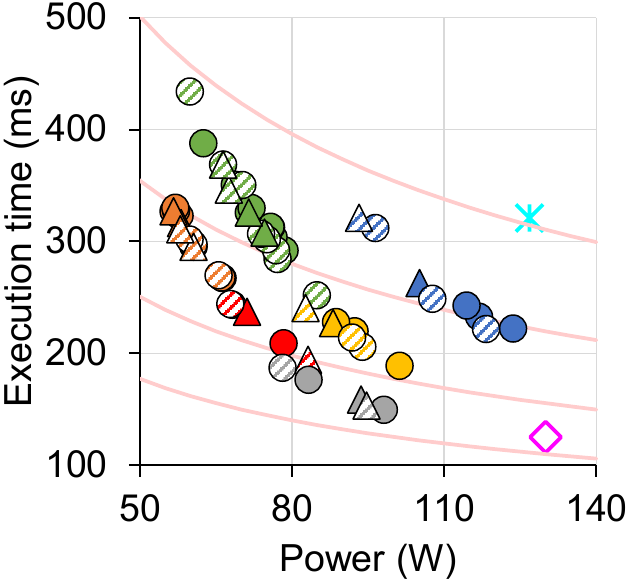}
  \caption{\resnet}
  \label{fig:pareto-resnet}
\end{subfigure}
\begin{subfigure}{.182\textwidth}
  \centering
  \includegraphics[height=1.17in]{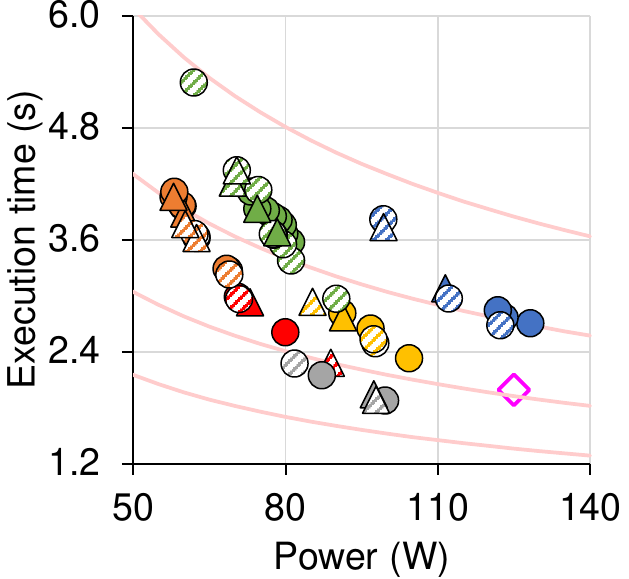}
  \caption{\sorting}
  \label{fig:pareto-sorting}
\end{subfigure}
\begin{subfigure}{.182\textwidth}
  \centering
  \includegraphics[height=1.17in]{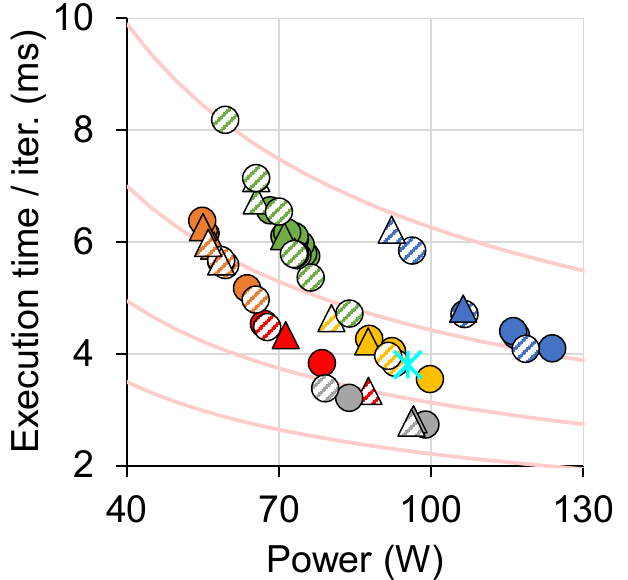}
  \caption{\helrsmall}
  \label{fig:pareto-helr256}
\end{subfigure}
\begin{subfigure}{.242\textwidth}
  \centering
  \includegraphics[height=1.17in]{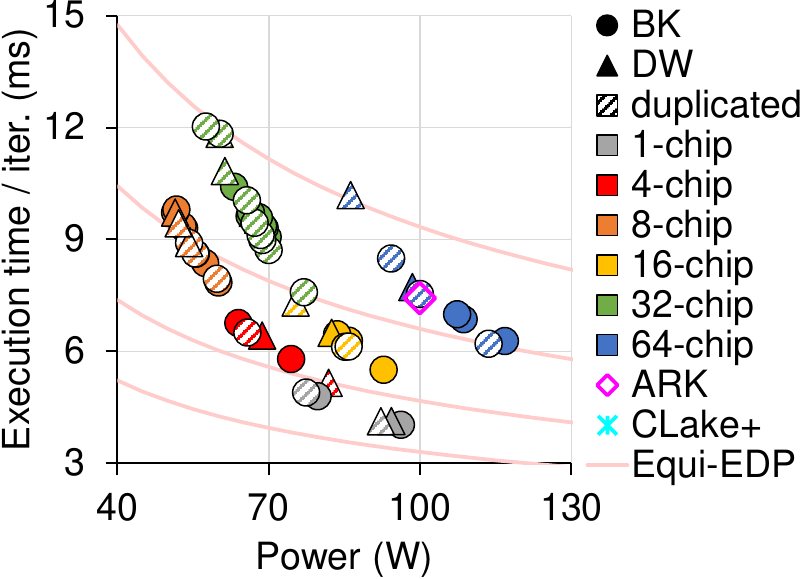}
  \caption{\helrlarge}
  \label{fig:pareto-helr1024}
\end{subfigure}
\caption{Performance comparison between \NAME default configurations and prior vector FHE accelerators, \subref{fig:pareto-boot}\subref{fig:pareto-resnet}\subref{fig:pareto-helr256} CLake+ and \subref{fig:pareto-boot}\subref{fig:pareto-resnet}\subref{fig:pareto-sorting}\subref{fig:pareto-helr1024} ARK, for the workloads. $n$-chip denotes the default configuration of \NAME with $n$ core chiplets, except for 1-chip, which is modified from 4-chip to integrate 4 cores into a monolithic die and has double bisection bandwidth to account for the bandwidth gap between NoC and NoP. BK: block clustering. DW: dimension-wise clustering. duplicated: using limb duplication.}
\label{fig:pareto}
\vspace{-0.13in}
\end{figure*}

\begin{figure}[t]
    \centering
    \includegraphics[width=0.95\columnwidth]{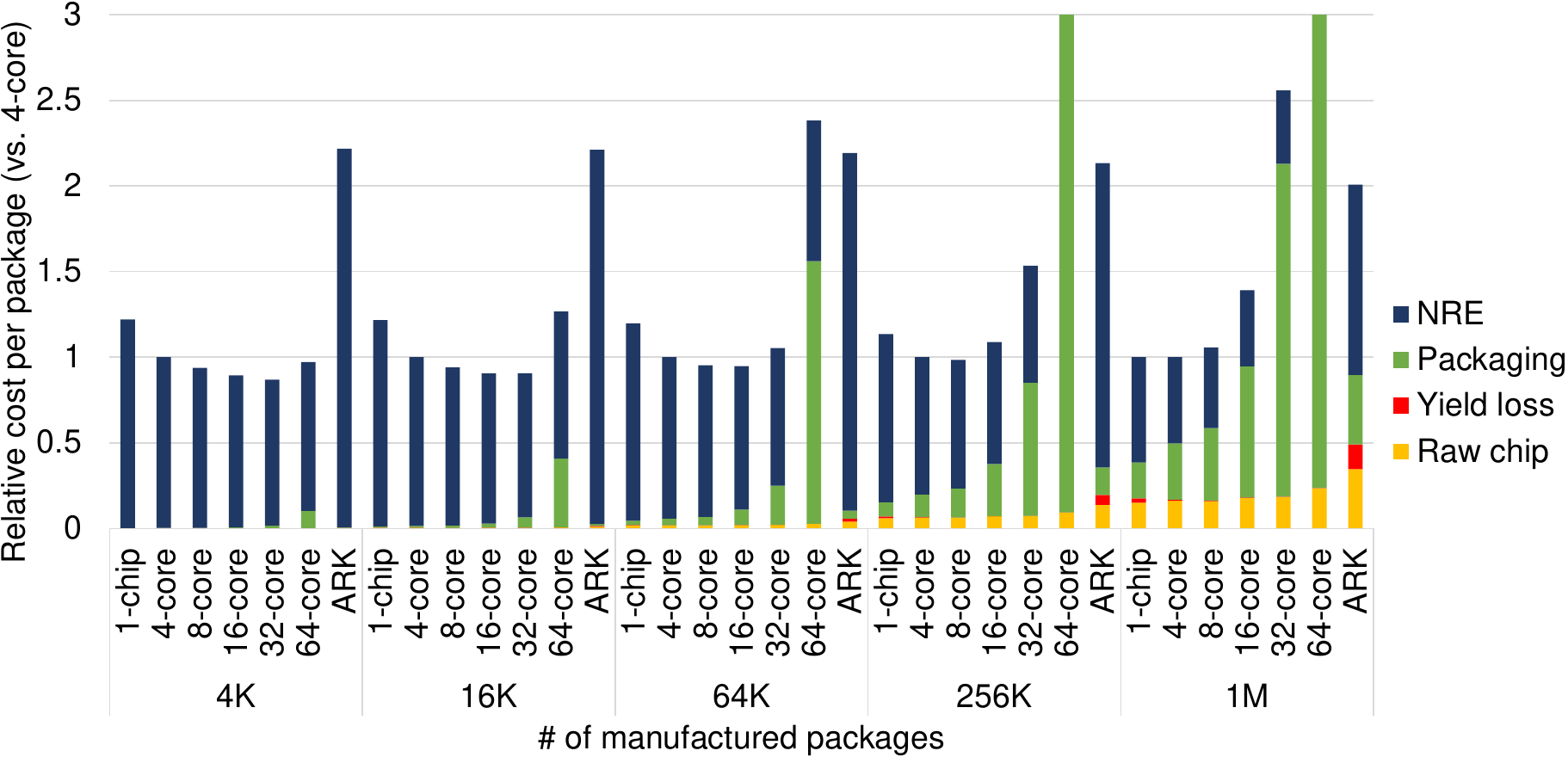}
    \caption{Relative manufacturing cost of \NAME and ARK systems compared to the 4-core \NAME system, depending on the number of manufactured systems.}
    \label{fig:cost-analysis}
    \vspace{-0.13in}
\end{figure}

\subsection{Implementation \& Default Configurations}

\noindent \textbf{Performance modeling:} We evaluate \NAME by implementing a cycle-accurate simulator for multi-core systems.
It takes a sequence of HE ops as input and translates it into a flow graph of instructions.
It divides instructions into micro-instructions and maps the micro-instructions to chiplets according to the data mapping method.
The simulator tracks the status of data transmission, data dependency between micro-instructions, and virtual channel allocation status of the NoP to avoid structural hazards and functionality issues.
Micro-instructions are scheduled based on this tracking data.
Our simulator adopts the decoupled data orchestration of CraterLake to prefetch data from HBM using the static nature of HE ops.

\noindent \textbf{Hardware modeling:} We synthesized logic components in RTL using the ASAP7 7.5-track 7nm process design kit~\cite{mj-2016-asap7}.
Large SRAM components with a bank size larger than 2KB were evaluated using a memory modeling tool, FinCACTI~\cite{isvlsi-2014-fincacti}.
We modified FinCACTI to match the IEEE IRDS roadmap~\cite{whitepaper-2018-irds}, ASAP7, and other published data in 7nm nodes~\cite{isscc-2017-7nm-sram, isscc-2018-7nm-sram-euv, iedm-2017-gf7nm, iedm-2016-tsmc7nm, vlsit-2018-samsung, isca-2021-tpuv4i}.
We used published data for HBM~\cite{micro-2017-finegrainedDRAM} and PHYs~\cite{whitepaper-2022-ucie, isca-2021-tpuv4i}.
We integrated the results into the performance modeling tool to automatically derive power and die area.
Overhead due to long wires is also estimated in the tool.
All components run at 1GHz.

\noindent \textbf{Manufacturing cost analysis:} We also estimated the relative cost of a \NAME package using the cost modeling tool from \cite{dac-2022-chiplet-cost}, results of which are shown in Fig.~\ref{fig:cost-analysis}.

\noindent \textbf{Routing and NoP implementation:} We used XY routing and $5\times5$ virtual channel routers~\cite{tpds-1992-virtual-channel} for NoP communication, which is a simple yet effective solution for the highly balanced traffic pattern of \NAME.
Each input and output port has 4 virtual channels.
We modeled NoP performance and energy using the specification of the UCIe advanced package with a PHY supporting 16GT/s of transfer rate~\cite{whitepaper-2022-ucie}.

\noindent \textbf{Memory instances:}
Two types of RFs, the main \emph{scratchpad RF} and an \emph{auxiliary RF} for key-switching, are utilized in \NAME.
The RFs together can provide 6 reads and 6 writes per lane in a cycle through bank interleaving~\cite{tc-1985-interleave}.
We modeled the NoP traffic for HBM access by injecting packets through edges connecting the I/O die and its adjacent cores on our simulator.
The bandwidth of this edge is the same as that of edges connecting the cores.
Two HBM stacks are utilized, each providing 500GB/s of bandwidth~\cite{jedec-2022-hbm3}.

\noindent \textbf{\NAME default configurations:}
The default configurations have different numbers of cores but similar levels of total computational throughput and memory bandwidth.
We can adjust the number of vector lanes in a core from 16 to 256.
We fixed $(\text{\# of cores})\times(\text{\# of lanes in a core})$ to 1,024 to make different configurations have similar total computational throughput.
We fixed the bisection bandwidth of the package to 2TB/s and decided the bandwidth of each edge by dividing it by the number of edges crossing the bisection.
The aggregate capacity of auxiliary RFs was set to 16MB (e.g., 1MB per core in a 16-core configuration).
Also, we used a total of 256MB for the scratchpad RF, which is sufficient for \NAME 
and is the same as the RF capacity of CraterLake.
Table~\ref{tab:area} shows the area breakdown of the default configurations of \NAME.

\subsection{Experimental Setup}

We utilized four representative FHE workloads that have been utilized for evaluating the previous accelerators.

\noindent \textbf{CKKS bootstrapping (\bootstrap)}:
FHE CKKS workloads require frequent bootstrapping, each involving hundreds of HE ops.
We performed bootstrapping for a ciphertext containing $2^{15}$ complex numbers.
We divided the execution time with the number of rescaling ops between consecutive bootstrapping execution (nine in our parameters with $L=48$) to account for the frequency of bootstrapping.

\noindent \textbf{CNN inference (\resnet)}:
We used a CKKS implementation~\cite{icml-2022-resnet} of the ResNet-20 model for CIFAR-10.
We report the latency of a single-image inference, which takes 2,271 seconds in the original single-threaded CPU implementation.

\noindent \textbf{Sorting $2^{14}$ numbers (\sorting)}:
We used the two-way sorting implementation from \cite{tifs-2021-sorting}.
The original 32-thread CPU implementation takes 23,066 seconds to finish.

\noindent \textbf{Logistic regression training (\helr)}: 
We used the implementation from \cite{aaai-2019-helr}. We performed encrypted training with binary classification for classifying the numbers $3$ and $8$ in the MNIST dataset. We iterated single-batch training 32 times and report the average execution time per iteration. We tested for two batch sizes of 256 (\helrsmall) and 1,024 (\helrlarge).

We compare \NAME with CraterLake and ARK using their reported performance and power values under 128-bit security constraints.
For 12/14nm CraterLake, optimistic compensations for the difference in the technology node (14nm to 7nm~\cite{iedm-2017-gf7nm}) were applied, which we refer to as CLake+.

\begin{figure*}[t]
    \centering
    \begin{subfigure}{0.307\textwidth}
    \centering
    \includegraphics[height=1.65in]{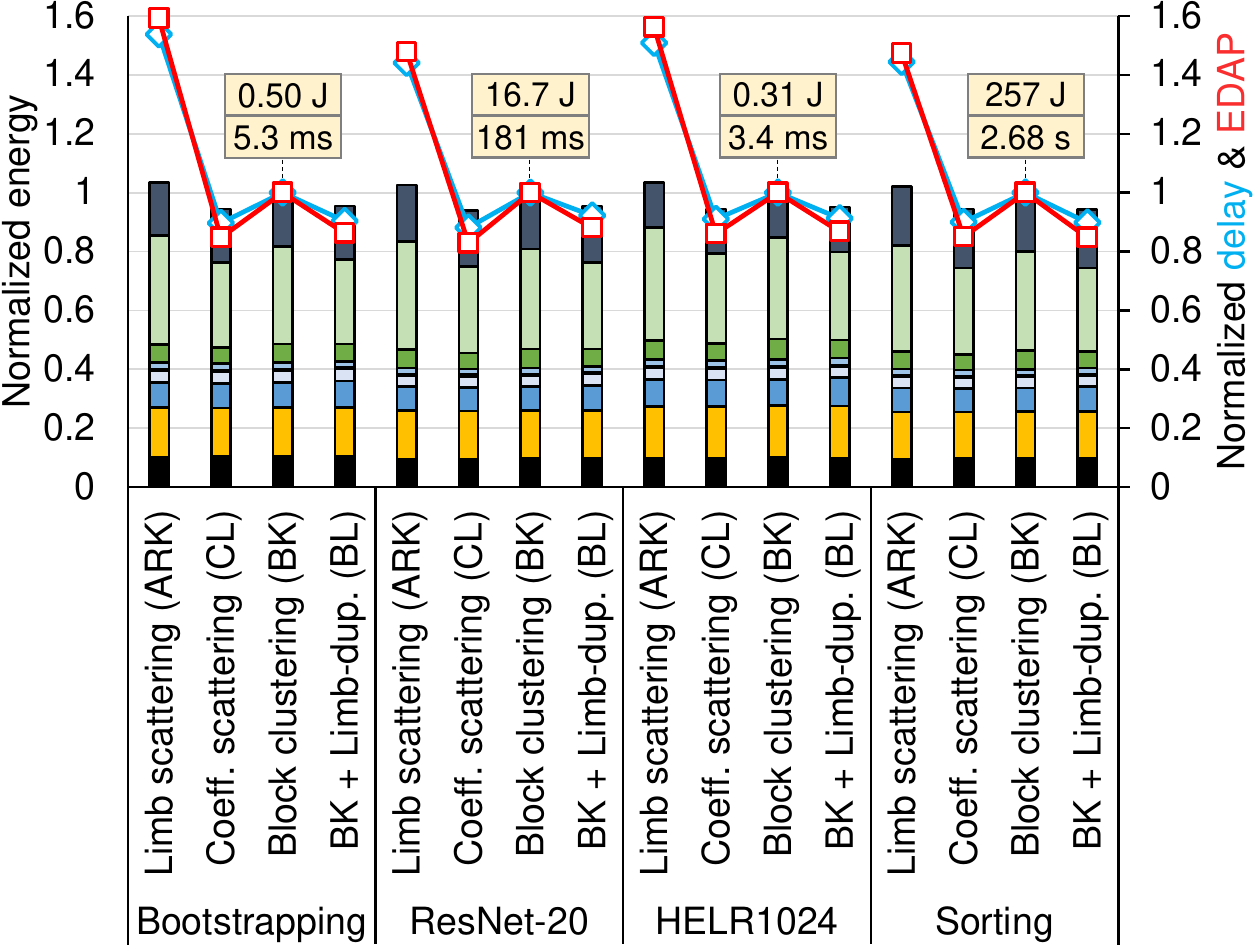}
    \caption{4$\times$4-BK-2$\times$2 configuration}
    \label{fig:ablation-4x4}
    \end{subfigure}
    \begin{subfigure}{0.365\textwidth}
    \centering
    \includegraphics[height=1.65in]{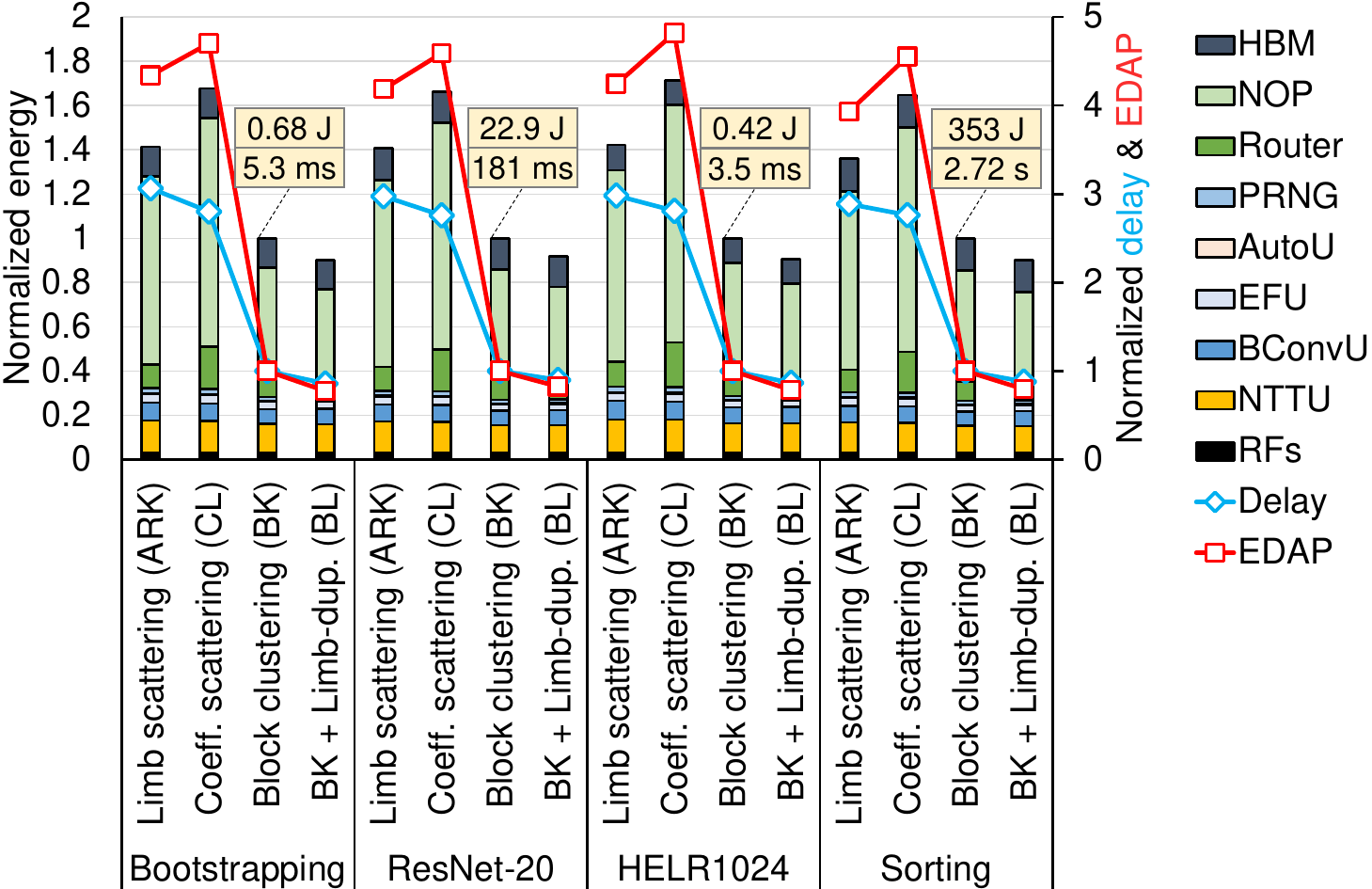}
    \caption{8$\times$8-BK-4$\times$4 configuration}
    \label{fig:ablation-8x8}
    \end{subfigure}
    \begin{subfigure}{0.307\textwidth}
    \centering
    \includegraphics[height=1.65in]{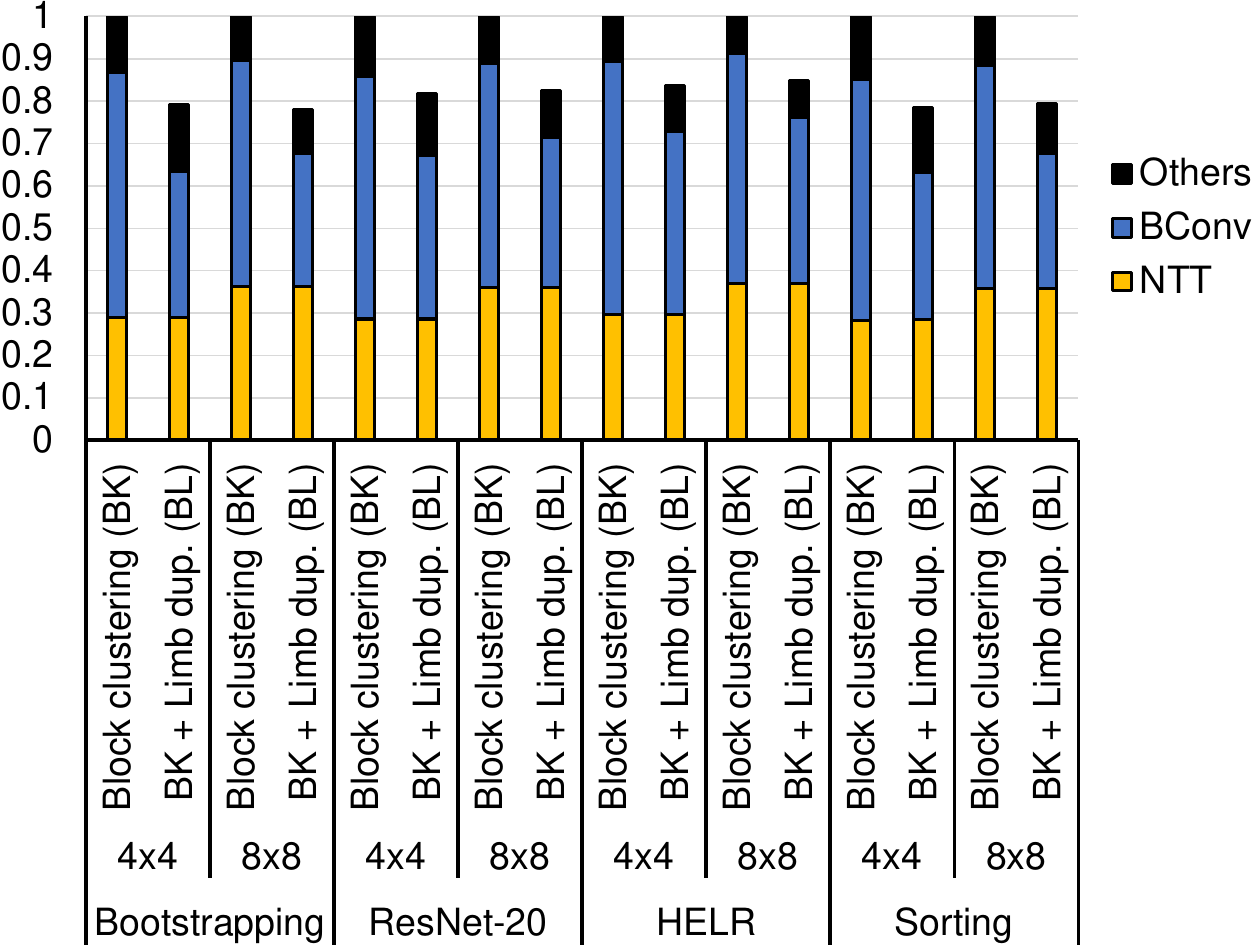}
    \caption{Data movement amount}
    \label{fig:limb_duplication}
    \end{subfigure}
    \caption{\subref{fig:ablation-4x4}\subref{fig:ablation-8x8} Delay, energy breakdown, and energy-delay-area product (EDAP)~\cite{micro-2009-mcpat} of \NAME when using the data mapping of ARK, CraterLake (CL), and our block clustering without (BK) and with limb duplication (BL). \subref{fig:limb_duplication} Amount of data movement between cores as \NAME performs each workload with and without limb duplication.} 
    \label{fig:ablation_study}
    \vspace{-0.05in}
\end{figure*}

\subsection{Performance \& Efficiency}

Among the default configurations, 4-core \NAME using dimension-wise clustering and limb duplication showed the best energy-delay product (EDP), whose values are shown with the 1$\times$ equi-EDP lines in Fig.~\ref{fig:pareto}.
We limited the number of limb clusters to eight to avoid severe fragmentation issues.

When we integrated the 4 cores into a single monolithic die and connected the cores in the NoC with 2$\times$ higher bisection bandwidth (1-chip in Fig.~\ref{fig:pareto}), additional performance and EDP enhancements were achieved.
1-chip \NAME resembles ARK with datapath reduced to 32 bits and showed a similar EDP with the original 64-bit ARK except for \helrlarge. ARK performs inferior in \helrlarge because we reordered the HE ops to further enhance the reuse of \evks in the memory-bound region, which ARK identifies as a bottleneck of \helr.

However, the fabrication cost of 1-chip \NAME or ARK is significatly higher than chiplet-based \NAME for reasonable production volumes (see Fig.~\ref{fig:cost-analysis}), mainly due to the high non-recurring engineering (NRE) cost stemming from the large chip size.
1-chip \NAME (respecitvely, ARK) has 1.22$\times$ (2.22$\times$) and 1.37$\times$ (2.48$\times$) higher NRE compared to 4-core and 16-core \NAME, 
Although NRE can be amortized for large production volumes, the cost of 1-chip \NAME becomes on par with that of 4-core \NAME when one million packages are produced, which is a huge amount for domain-specific accelerators.
Meanwhile, ARK is always expensive than 4-core \NAME due to higher raw chip fabrication cost and yield loss.

\setlength{\tabcolsep}{2.1pt}
\begin{table}[t]
\centering
\caption{Execution time and relative energy-delay-area product (EDAP)~\cite{micro-2009-mcpat} of \NAME and prior vector FHE accelerators.}
\label{tab:comparison}
\small
\begin{tabularx}{0.95\columnwidth}{l|LLLLL}
\toprule
\multicolumn{6}{l}{\textbf{Execution time (ms)}, the lower the better}\\
\midrule
 & \multirow{2}{*}{CLake+} & \multirow{2}{*}{ARK} & \multicolumn{3}{l}{CiFHER} \\
& & & 4 cores & 16 cores & 64 cores \\ 
\midrule
 \bootstrap & 0.79 & \textbf{0.44} & 0.62 & 0.64 & 0.73 \\
 \resnet & 321 & \textbf{125} & 194 & 189 & 222\\
 \sorting & - & \textbf{1990} & 2282 & 2328 & 2683 \\
 \helrsmall & 3.81 & - & \textbf{3.34} & 3.55 & 4.09 \\
 \helrlarge & - & 7.42 & \textbf{5.16} & 5.50 & 6.20 \\
  \midrule
  \multicolumn{6}{l}{\textbf{Relative EDAP (vs. 4-core \NAME)}, the lower the better }\\
  \midrule
   \bootstrap & 2.04$\times$ & 1.46$\times$ & 1.00$\times$ & 1.35$\times$ & 2.63$\times$ \\
   \resnet & 4.09$\times$ & 1.20$\times$ & 1.00$\times$ & 1.26$\times$ & 2.56$\times$ \\
  \sorting & - & 2.04$\times$ & 1.00$\times$ & 1.34$\times$ & 2.62$\times$\\
   \helrsmall & 1.40$\times$ & - & 1.00$\times$ & 1.41$\times$ & 2.80$\times$ \\
    \helrlarge & - & 4.70$\times$ & 1.00$\times$ & 1.42$\times$ & 2.77$\times$\\
 \bottomrule
\end{tabularx}
\vspace{-0.13in}
\end{table}
\setlength{\tabcolsep}{6pt}

Under the fixed computational throughput and bisection bandwidth of the package, as we split the chip and utilize smaller cores, performance and efficiency inevitably decline because more NoP data transfer is required.
Thus, the area constraint is a decisive factor in an MCM FHE accelerator design.
If the cost budget allows, as in recent AMD server CPUs (74mm\textsuperscript{2} per compute die~\cite{isscc-2020-amdchiplet}), utilizing four cores would be best.
Table~\ref{tab:comparison} shows that 4-core \NAME enhances efficiency in terms of energy-delay-area product (EDAP)~\cite{micro-2009-mcpat} with comparable performance to prior monolithic FHE accelerators due to significantly lower power dissipation (see Fig.~\ref{fig:pareto}).
Compared to ARK, 4-core \NAME performs 1.15$\times$
slower but reduces EDAP by 2.03$\times$
in geometric mean (geomean) of the workloads.
Compared to CLake+, it performs 1.34$\times$ faster and results in a 2.27$\times$ EDAP reduction.
By contrast, if we can only afford dies as small as recent domain-specific MCM accelerators, such as NN-Baton~\cite{isca-2021-nnbaton} (2mm\textsuperscript{2}) and Simba~\cite{micro-2019-simba} (6mm\textsuperscript{2}), 16-core or 64-core \NAME needs to be utilized.

Our data mapping and limb duplication minimize the performance and efficiency degradation of many-core configurations.
The best 16-core and 64-core configurations respectively perform only 1.03$\times$ and 1.18$\times$ slower in geomean than 4-core.
Compared to 4-core, EDP increases to 1.23$\times$ (16-core) and 1.94$\times$ (64 core), and EDAP increases to 1.35$\times$ (16-core) and 2.67$\times$ (64-core) in geomean using our optimized configurations, whereas a na\"{i}ve configuration (e.g., 64-core using limb scattering with ARK's method) would require 17.1$\times$ higher EDP and 23.6$\times$ higher EDAP in bootstrapping.

Meanwhile, due to the skewed organization, 8-core (2 × 4) and 32-core (4 × 8) configurations suffer from an imbalance in NoP throughput between the horizontal and vertical directions, increasing the tail latency.
As a result, performance drops severely, making these configurations less attractive.

\begin{figure*}[t]
    \centering
    \begin{subfigure}{.31\textwidth}
        \centering
        \includegraphics[width=\linewidth]{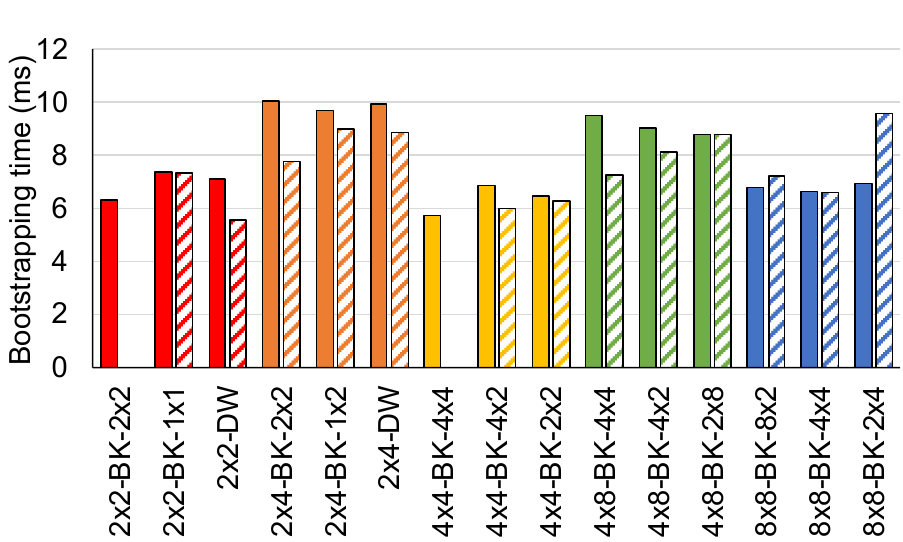}
        \caption{\NAME baseline}
        \label{fig:selective-default}
    \end{subfigure}
    \begin{subfigure}{.31\textwidth}
        \centering
        \includegraphics[width=\linewidth]{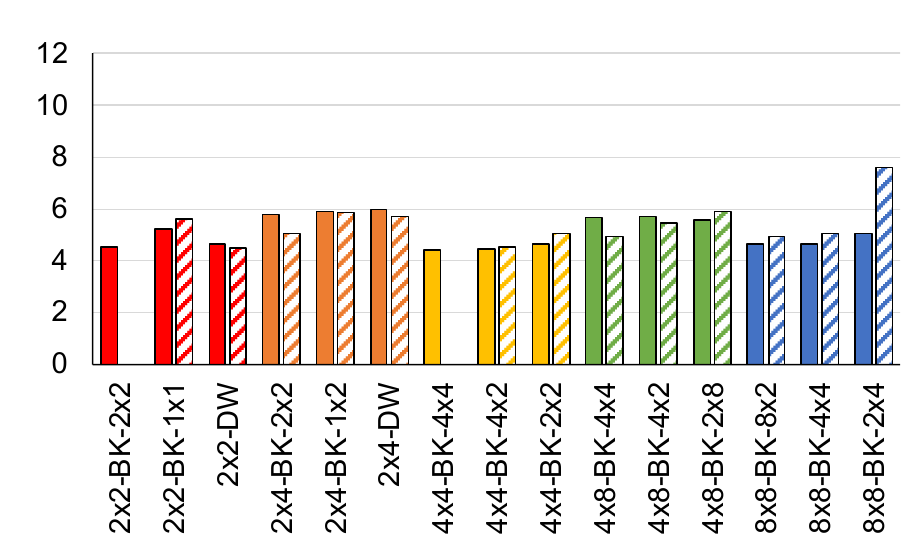}
        \caption{\NAME with double NoP bandwidth}
        \label{fig:selective-2xnop}
    \end{subfigure}
    \begin{subfigure}{.31\textwidth}
        \centering
        \includegraphics[width=\linewidth]{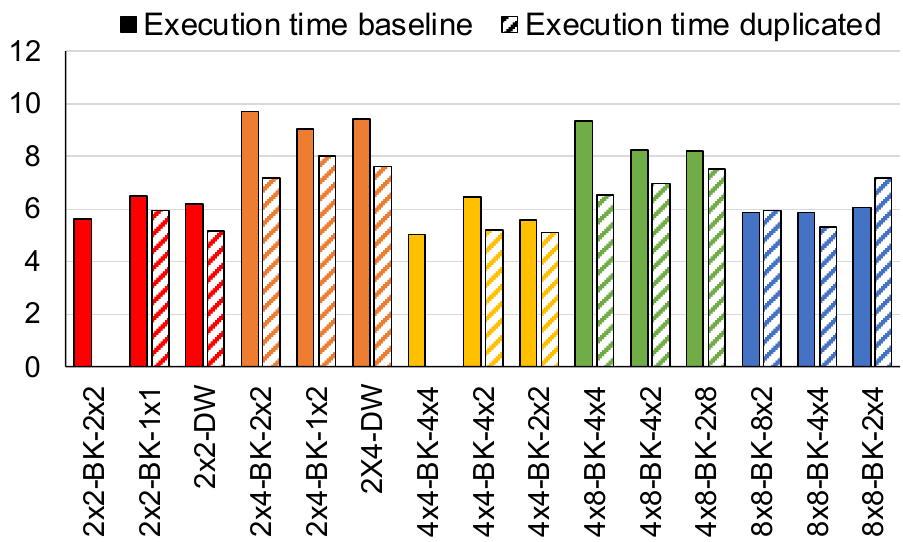}
        \caption{\NAME with double compute}
        \label{fig:selective-2xcomp}
    \end{subfigure}
    \caption{Bootstrapping performance of \NAME with/without limb duplication. The three best-performing mapping configurations were selected for each number of cores. ARK's redistribution method for BConv was used when limb duplication was not used.}
    \label{fig:selective_output}
    \vspace{-0.13in}
\end{figure*}

\subsection{Sensitivity Study}

We estimated energy, delay, and EDAP of \NAME while incrementally applying block clustering (4$\times$4-BK-2$\times$2 and 8$\times$8-BK-4$\times$4) and limb duplication (see Fig.~\ref{fig:ablation_study}).
Limb scattering of ARK shows 1.44--1.53$\times$ (4$\times$4) and 2.89--3.06$\times$ (8$\times$8) slow execution time compared to block clustering.
Coefficient scattering of CraterLake is 1.10--1.13$\times$ faster than our method for the 4$\times$4 configuration, whereas it is 2.76--2.81$\times$ slower for 8$\times$8.
The data exchange pattern during limb scattering is not all-to-all as we can group some limbs together to have data transfer occur within each group.
By contrast, coefficient scattering requires all-to-all data exchange and thus scales inferior to 
limb scattering.
Coefficient scattering consumes 2.07$\times$ more energy on NoP and shows 4.66$\times$ higher (worse) EDAP than our mapping in geomean for 8$\times$8.
Yet, all-to-all data exchange between 16 cores is manageable, making coefficient scattering faster than our mapping for 4$\times$4.

Fig.~\ref{fig:limb_duplication} compares the amounts of data movement between the cores of \NAME, with and without limb duplication.
Limb duplication consistently eliminated 18--22\% of the data movement between the cores.
When applying limb duplication to our mapping in the 4$\times$4 configuration, we obtained 1.10$\times$ delay and 1.16$\times$ EDAP reductions in geomean, and the delay and EDAP gaps with coefficient scattering were also reduced to only 1.3\% and 2.1\% in geomean.

\begin{figure}
\centering
\begin{subfigure}{0.42625\columnwidth}
  \centering
  \includegraphics[height=1.6in]{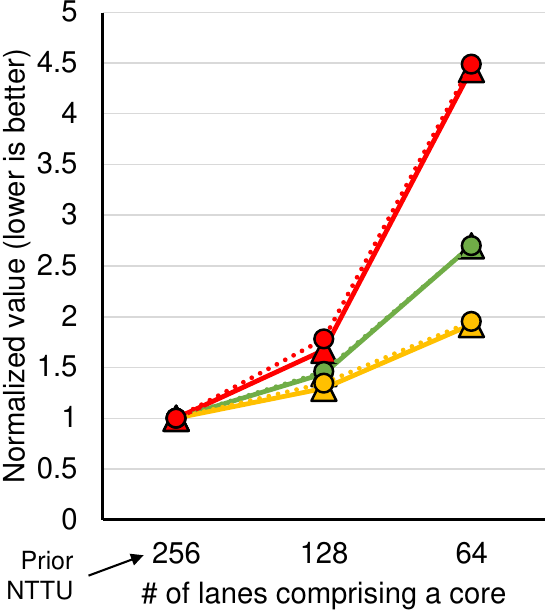}
  \caption{2x2-DW}
  \label{fig:eval-composable-2x2}
\end{subfigure}%
\begin{subfigure}{0.56375\columnwidth}
  \centering
  \includegraphics[height=1.6in]{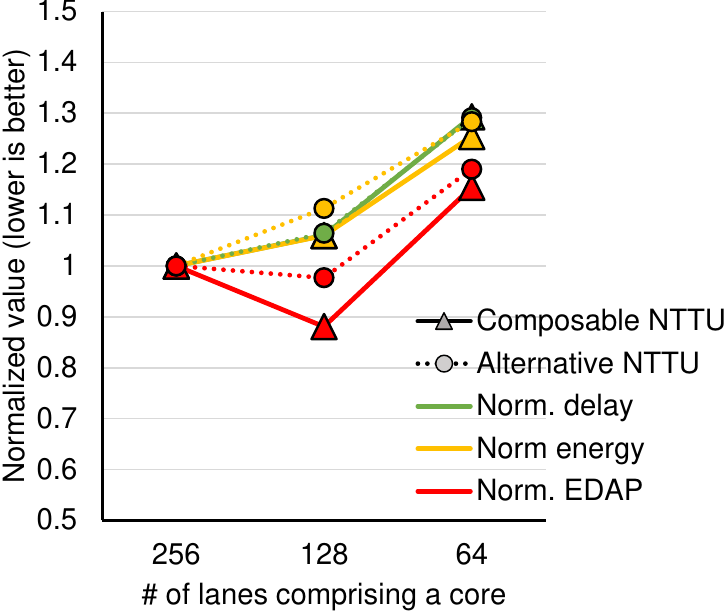}
  \caption{4x4-BK-2x2}
  \label{fig:eval-composable-4x4}
\end{subfigure}
\caption{Normalized delay, energy, and EDAP of CiFHER when performing bootstrapping with various single-core computational throughput using the composable NTTU or an alternative NTTU.}
\label{fig:eval-composable}
\vspace{-0.13in}
\end{figure}

To assess the efficiency of composable NTTUs, we also measured delay, energy, and EDAP of \NAME during bootstrapping with different single-core computational throughput, represented by the number of lanes.
We either use composable NTTUs or alternative NTTUs, which extend upon prior NTTUs by interpreting a length-$N$ polynomial as a three-dimensional vector ($N \!=\! \text{\#}_{lane} \times \text{\#}_{lane} \times N_{rest}$) and performing a sequence of $\text{\#}_{lane}$-point, $\text{\#}_{lane}$-point, and $N_{rest}$-point NTTs.
Similar to prior NTTUs, twisting and transposing steps are appropriately inserted in between NTT steps.
Alternative NTTUs can also adjust the computational throughput according to the number of lanes ($\text{\#}_{lane}$); however, whereas data exchange in composable NTTUs mostly occur in between adjacent lanes, alternative NTTUs require data exchange with distant lanes, increasing overall energy and area due to long wires.

Composable NTTUs increase the efficiency of \NAME for many-core configurations by enabling a compact MCM architecture with a balance between computational throughput and available data movement bandwidth. 
In the 2x2-DW configuration, where data transfer through NoP is relatively less heavy and high computational throughput is required for each core, energy and delay increased when prior NTTUs were replaced with composable NTTUs with lower throughput (see Fig.~\ref{fig:eval-composable-2x2}).
Specifically, the delay (respectively, energy) increases by $1.43\times (1.29\times)$ and $2.7\times (1.92\times)$ as the number of lanes in a core decreases from 256 to 128 and 64 by utilizing composable NTTUs.
Conversely, in the 4x4-BK-2x2 configuration that demands substantial NoP data transfer, halving the number of lanes by using composable NTTUs do not increase the delay by a large margin.
Using 128 lanes leads to a slight 6\% increase in delay and improves the overall EDAP by 12\% due to smaller chiplet area.

Composable NTTUs are more efficient in terms of energy and area compared to alternative NTTUs.
In Fig.~\ref{fig:eval-composable-4x4}, the reduction to 128 lanes using alternative NTTU also leads to an around 2\% enhancement in EDAP. Nevertheless, due to the substantial wire requirements of alternative NTTUs, the energy consumption and area footprint are larger by $1.05\times$ for both when compared to those of composable NTTUs, culminating in an EDAP value $1.10\times$ larger than that achieved with composable NTTUs.

\subsection{Limb Duplication Effects}

The effects of limb duplication differ by each configuration.
We selected the three best-performing configurations for each number of cores in Fig.~\ref{fig:pareto-boot} and compared the performance of bootstrapping with and without applying limb duplication.
Fig.~\ref{fig:selective-default} shows the results.
Limb duplication improved bootstrapping performance by up to 31\% for 4$\times$8-BK-4$\times$4 but rather caused 28\% performance degradation for 8$\times$8-BK-2$\times$4.

Although limb duplication makes fewer limbs exchanged, the bursty nature of data broadcasting causes network contention and introduces additional latency.
The overhead of broadcasting in limb duplication is affected by the arrangement of limb clusters.
Only 8$\times$8-BK-2$\times$4 in Fig.~\ref{fig:selective-default} has a coefficient cluster composed of eight cores; coefficient clusters in the other configurations consist of one to four cores.
The large number of cores in a coefficient cluster greatly increases the broadcasting latency, degrading performance.

As limb duplication is effective for reducing the NoP pressure, increasing the NoP bandwidth reduced the performance gains from limb duplication (see Fig.~\ref{fig:selective-2xnop}).
Limb duplication enhanced the performance of the baseline configurations by 7\% in geomean.
With the NoP bandwidth doubled, limb duplication rather decreased performance by 3\% in geomean, and the maximum performance gain observed for 4×8-BK-4×4 was reduced to 14\% from 31\% in the baseline.

When we intensified the NoP bandwidth bottleneck by doubling computational throughput, performance gains from using limb duplication increased (see Fig.~\ref{fig:selective-2xcomp}).
In these configurations, limb duplication enhanced performance by 14\% in geomean, and the maximum performance gain reached 43\%.

\subsection{Scalability}

\begin{figure}[t]
    \centering
    \includegraphics[width=0.99\columnwidth]{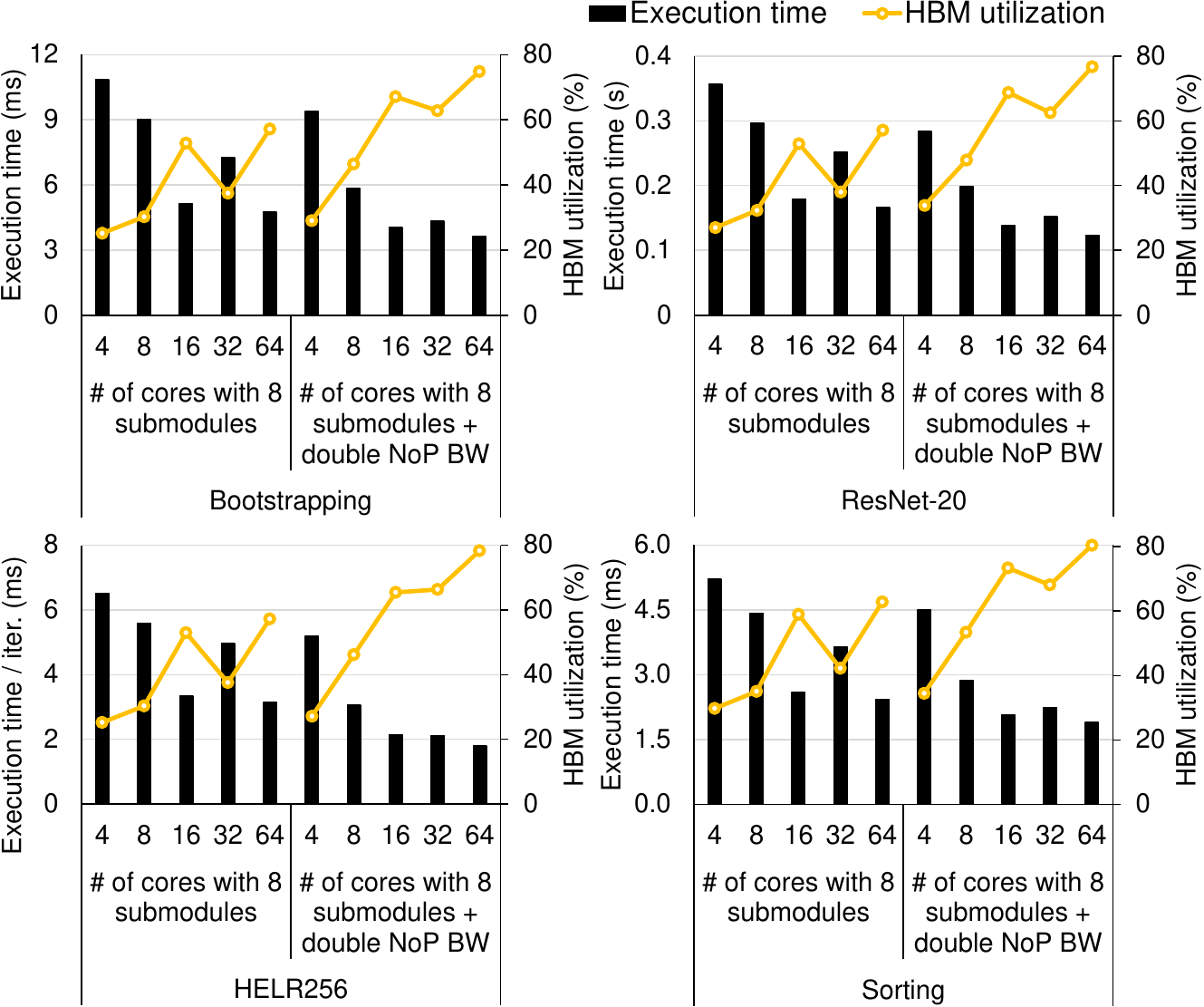}
    \caption{Performance of \NAME while scaling the number of cores under default and double NoP bandwidth conditions. 
    The number of NTT submodules in each core is fixed to eight.
    }
    \label{fig:scalability}
    \vspace{-0.13in}
\end{figure}

To evaluate the scalability of \NAME, we estimated the performance of each workload while increasing the number of cores in Fig.~\ref{fig:scalability}.
We set the internals of a core fixed to an eight-NTTU-submodule setting (128 lanes).
Limb duplication and $d_x \times d_y$-BK-$\sfrac{d_x}{2} \times \sfrac{d_y}{2}$ block clustering were used for a $d_x\times d_y$ package configuration.
We fixed the HBM bandwidth and the bisection bandwidth of a package.
Thus, the bandwidth of a network edge decreases as the number of cores increases.

The limited bandwidth of NoP restricts the scalability of \NAME.
When we increased the total computational throughput by increasing the number of cores from 4 to 16, 1.99--2.11$\times$ of speedups were achieved for the workloads.
Increasing from 16 to 64 only resulted in a 1.07$\times$ speedup. 
Therefore, increasing the number of cores without consideration for proper sizing of the core has an adversarial effect on efficiency.
Meanwhile, skewed configurations (8-core and 32-core), which already suffer from the NoP bandwidth imbalance, showed suboptimal performance.

HBM bandwidth becomes the next limiting factor for scalability when NoP bandwidth is sufficient.
Doubling NoP bandwidth mitigates the NoP pressure and improves the performance of the 16-core configuration by 1.34$\times$ in geomean.
Scalability slightly improves as 4-to-16 and 16-to-64 number-of-cores transitions result in 2.23$\times$ and 1.13$\times$ speedups.
However, the degree of speedups still falls largely behind the increase in the number of cores as the hardware becomes bottlenecked by the HBM bandwidth, whose utilization reaches 75--80\% in the 64-core configuration.
\section{Related Work}
\label{sec:related}


\noindent \textbf{Function-level acceleration and leveled HE (LHE):}
\cite{access-2021-demystify} accelerates non-RNS CKKS using Intel AVX-512 instructions and GPUs.
HEXL~\cite{wahc-2021-hexl} specifically utilizes AVX512-IFMA52 instructions for further CPU optimization.
\cite{ipdps-2022-intelgpu} accelerates (i)NTT through high-radix FFT and instruction-level optimizations using Intel GPUs.
\cite{iiswc-2020-ntt} identifies off-chip memory access as the performance bottleneck of (i)NTT on GPUs and applies a better use of the memory hierarchy of GPUs, improving data reusability to address the bottleneck.
\cite{al2018high} accelerates the BFV~\cite{siam-2014-bfv1, tetc-2019-bfv} scheme by replacing the major functions with ones more suitable for GPUs.
\cite{tpds-2021-multigpu} further partitions data to multiple GPUs to exploit abundant parallelism inside BFV.

Numerous FPGA works attempted to optimize core HE ops by deploying tailored datapaths.
\cite{hpca-2019-roy} and \cite{tc-2020-heaws} proposed specialized FUs for the BFV scheme.
HEAX~\cite{asplos-2020-heax} and coxHE~\cite{date-2022-coxhe} designed FPGA modules targeting key-switching, a dominant HE op in CKKS.
\cite{fccm-2020-sunwoong-ntt} supports larger parameters adequate for FHE CKKS; however, it only accelerates (i)NTT.


\noindent \textbf{Torus LWE (TLWE) FHE:}
TLWE-based FHE schemes~\cite{jc-2020-tfhe, eurocrypt-2015-fhew} are fundamentally different from RLWE in that usually a single prime modulus (limb) is utilized.
Recent studies have specifically targeted the TFHE~\cite{jc-2020-tfhe} scheme.
cuFHE~\cite{github-cufhe} speeds up TFHE computations on a GPU by utilizing primitive functions accelerated in cuHE~\cite{balkans-2015-cuhe}, which uses a specialized prime to reduce the cost of modular reduction.
\cite{iccad-2022-xhec} implements an FPGA datapath for an optimized TFHE bootstrapping algorithm using inter-operation parallelism.
In TFHE, floating-point Fourier transform is often used instead of NTT.
\cite{host-2020-cpuandgpufhe} utilizes bit-level parallelism and the NVIDIA cuFFT library to accelerate TFHE on GPUs.
Matcha~\cite{dac-2022-matcha} is the first ASIC proposal and devises an approximate integer FFT method that replaces mult ops with shift and add ops.

\noindent \textbf{Ring LWE (RLWE) FHE:} 
\cite{tches-2021-100x} applies kernel fusion and parameter optimization to accelerate CKKS bootstrapping on GPUs.
TensorFHE~\cite{hpca-2023-tensorfhe} utilizes the tensor cores in recent NVIDIA GPUs for (i)NTT. \cite{micro-2023-GME} introduces new microarchitectural extensions that minimize redundant data access to DRAM and enhance the computational throughput of modular reduction operations, further boosting the performance of FHE operations on current GPUs.
FAB~\cite{hpca-2023-fab} reduces the working set to alleviate the off-chip memory bandwidth bottleneck even with limited FPGA memory space.
Poseidon~\cite{hpca-2023-poseidon} builds a 512-lane vector FPGA architecture with FUs reducing the on-chip memory bandwidth pressure, such as high-radix NTTUs.
GPU and FPGA solutions achieve orders of magnitude higher performance compared to CPU but recent ASIC FHE accelerators~\cite{micro-2021-f1, isca-2022-craterlake, isca-2022-bts, micro-2022-ark} introduced in Section~\ref{sec:background:prior} achieve even higher performance gains. \cite{aikata2023reed} also introduces an FHE accelerator design using chiplets to reduce chip size but only covered a fixed number of chiplets, set as the sweet spot.

\section{Conclusion} \label{sec:conclusion}
We have proposed \NAME, a fully homomorphic encryption (FHE) accelerator based on a multi-chip module (MCM).
We devised methods for minimizing die-to-die communication, which incurs serious latency and energy overhead for an MCM.
Our data mapping limits data communication to happen within a cluster, reducing the traveling distance of data and alleviating the network pressure.
A novel limb duplication algorithm eliminates data transfer caused by redistributing output during base conversion, a key function of FHE.
By utilizing multiple chiplets with a resizable structure, enabled by using composable number-theoretic transform units, we created various optimized MCM package configurations of \NAME.
Multiple chiplets, each configurable to be as small as 4.28mm\textsuperscript{2}, collaborate efficiently with our data mapping methods and algorithms; the resulting \NAME performs comparable to state-of-the-art monolithic ASIC accelerators with significantly reduced area and energy.

\section*{Acknowledgement}

This work was supported in part by Samsung Advanced Institute of Technology, Samsung Electronics Co., Ltd. and by Institute of Information \& communications Technology Planning \& Evaluation (IITP) grant funded by the Korea government (MSIT) (No.2021-0-01343 and IITP-2023-RS-2023-00256081).
The EDA tool was supported by the IC Design Education Center (IDEC), Korea.
Jung Ho Ahn, the corresponding author, is with the Department of Intelligence and Information, the Interdisciplinary Program in Artificial Intelligence, and ICT (Institute of Computer Technology), Seoul National University, South Korea.


\balance
\bibliographystyle{IEEEtranS}
\bibliography{refs}

\end{document}